\documentclass[hyper(*)]{JHEP3}

\usepackage{epsfig,multicol,bbm}
\usepackage{enumerate}
\usepackage{bibentry}

\title{Network, Cluster coordinates and $\mathcal{N}=2$ theory I}

\author{Dan Xie

\\ School of Natural Sciences, Institute for Advanced Study \\
Princeton, NJ 08540, USA}

\abstract{Combinatorial methods are developed to find the cluster coordinates for moduli space of flat connections 
which is describing the Coulomb branch of higher rank $\mathcal{N}=2$  theories derived by compactifying six dimensional $(2,0)$ theory on a punctured Riemann surface.
The construction starts with  a triangulation of the punctured Riemann surface and a  further tessellation of all the triangles. The tessellation is used to construct  a bipartite network
from which a quiver can be read straightforwardly. We prove that the quivers is independent of the triangulations and cyclic choice of the punctures if all the puncture has at most
one column with height larger than one.
These coordinates are important in studying 
BPS wall crossing, line operators, and surface operators of these theories; and they are also useful in exploring three dimensional Chern-Simons theory and the corresponding $\mathcal{N}=2$ gauge
theory,  integrable systems, etc.}

\begin{document}
\newpage
\section{Introduction}
There are several exciting developments in understanding dynamics of quantum field theory in last couple of years. 
A large class  of new 4d $\mathcal{N}=2$ superconformal field theories (SCFT) (we call them theories of class ${\cal S}$)
 are engineered using the six dimensional $(2,0)$ theory \cite{Gaiotto:2009we} generalizing the remarkable observation on S duality properties of $\mathcal{N}=2$ SCFT \cite{Argyres:2007cn}.
 New exact quantities like stable BPS spectrum \cite{Gaiotto:2008cd,Gaiotto:2009hg, Gaiotto:2010be, Gaiotto:2011tf, Cecotti:2011rv, Alim:2011ae, Alim:2011kw} and partition function on compact manifold
are successfully calculated \cite{Pestun:2007rz,Kapustin:2009kz}. There are also unexpected identities between observables of theories in different space-time 
dimensions: It is conjectured that the partition function of 4d $\mathcal{N}=2$ SU(2)
generalized quiver gauge theory on $S^4$ is identical to the correlation function of two dimensional Liouville theory \cite{Alday:2009aq};
It is shown that the Nekrasov partition function of 4d theory on Omega deformed manifold is related to the Yang-Yang function of the  
quantum integrable system involving in the Seiberg-Witten solution of 4d theory \cite{Nekrasov:2009uh,Nekrasov:2009ui,Nekrasov:2009rc};
the duality between various physical quantities of Chern-Simons theory,
3d  $\mathcal{N}=2$ gauge theory and 2d Liouville theory are also discovered \cite{Dimofte:2011ju,Dimofte:2011py,Terashima:2011qi}. 

Moduli space of complex flat connections $\mathcal{M}$ on a genus $g$ Riemann surface with $n$ punctures \cite{Hitchin:1987ab,Hitchin:1987bc}
and its hyperkahler extension shows up in all these physical theories. For example,
 it is the Coulomb branch of the four dimensional ${\mathcal N}=2$ theory compactified 
on a circle; the Seiberg-Witten  integrable system is described in another complex structure of the hyperkahler extension;
it is  the phase space of the complex Chern-Simons theory on a three manifold; the moduli space of the corresponding 3d $\mathcal{N}=2$ theory is 
described by one of its Lagrangian submanifolds;
one of the real slices of $\mathcal{M}$ is the (higher) Teichmuller space and whose quantization is isomorphic 
to (Toda) Liouville  theory, etc.

A special coordinate system called cluster coordinates on $\mathcal{M}$ play an extremely important role in many recent developments.
The cluster coordinates are described by a collection of triples $(\epsilon_{ij}, X_i, Y_i)$, where $\epsilon_{ij}$ is an antisymmetric tensor which can 
be represented by a quiver. Different triples are related by mutation which acts like Seiberg duality on the corresponding quiver.
There are many remarkable properties of these coordinates  and more details can be found in \cite{Fomin2001}.  For SU(2) group,  each triple of the cluster 
coordinates are described by a triangulation of the punctured Riemann surface, from which a tensor $\epsilon_{ij}$ can be easily read.
The mutation on triples is nicely represented geometrically as the flip which relates different triangulations.
These coordinates are extensively used in the following applications:

1.  Finding the BPS spectrum and studying the Wall crossing behavior of a  4d $\mathcal{N}=2$ theory \cite{Gaiotto:2008cd,Gaiotto:2009hg,Cecotti:2011rv, Alim:2011ae, Alim:2011kw}.

2. The classification of the line operators \cite{Drukker:2009tz}, the BPS wall crossing in the presence of line defects and surface defects \cite{Gaiotto:2010be, Gaiotto:2011tf}.

3. The quantization of Teichmuller theory and equivalently the Liouville theory \cite{chekhov-1999-120}. 

4. The quantization of complex Chern-Simons theory on hyperbolic three manifolds \cite{Dimofte:2011gm,Dimofte:2011jd,Nagao:2011aa}, and finding  a corresponding
3d $\mathcal{N}=2$ theory from the same manifold \cite{Dimofte:2011ju}. 

.......

However, most of these studies are centered around SU(2) theory. The analog of cluster coordinates for the higher rank group with full
punctures are introduced in a seminal paper by Fock and Goncharov (FG) \cite{fock-2003}, though there is very few applications up to now. 
The generalization to the non-full puncture case is not available in the literature. It is urgent to find such cluster coordinates for non-full puncture  since they are very important 
in studying general ${\mathcal N}=2$ theories, for example, $\mathcal{N}=2$ theory with Lagrangrian description require non-full puncture.
The purpose of this paper is to fill this gap.

Our construction also starts with a triangulation of the punctured Riemann surface as used in  SU(2) theory.  Further 
tessellation for each triangle is done using the rule given in \cite{Benini:2009gi} which provides a brane construction for
three punctured  theories. A bipartite network is then introduced on each triangle based on the 
tessellation from which a quiver can be read (similar construction for the full puncture case is also proposed in \cite{Goncharov:2012aa}). 
The full network on the punctured Riemann surface is derived by gluing network of each triangle in a natural way and the 
full quiver can also be defined. Such type of network 
is appearing previously in studying the dimer and $\mathcal{N}=1$ gauge theories \cite{Franco:2005rj}, and it is used extensively recently in
studying scattering amplitude of $\mathcal{N}=4$ super Yang-Mills theory \cite{Nima:2012}. 

The construction is based on a triangulation and also a choice of closed path connecting punctures which defines a cyclic order 
of punctures. 
For a fixed closed path, there are many different triangulations which are related by the local move called flip, we 
prove that the quivers for different triangulations are related by quiver mutations with some restrictions on the punctures. 
Moreover, if there are more than one possible choices of closed paths, we prove that they define the same quiver if all the punctures have 
at most one column with height larger than one.

The quiver mutation allowed is special in the sense that we only do the mutation on quiver nodes with four edges. Interestingly, 
all the quiver nodes (with some doubling for the special case) for the SU(2) theory has this property. So if we confine ourselves 
to these special quiver mutations, the quiver is under good control and it will quickly come into a huge mess if 
we do arbitrary quiver mutations.  It is remarkable that once we restrain to these special quiver mutations, the cluster coordinates
for higher rank theory are exactly parallel as the SU(2) case.

We focus on constructing cluster coordinates for moduli space related to theories of class ${\cal S}$. There are parallel stories for the Argyres-Douglas (AD) theory \cite{Argyres:1995jj} and 
asymptotical free (AF) theory which require irregular singularities  to have a six dimensional construction. Such theories are studied 
in a sequel to this paper \cite{Xie:2012mm}.  More general theories can be engineered using toric geometry which one could find a dimer and 
a quiver, we expect these quivers to define cluster coordinates for the moduli space and it would be interesting to verify this in detail. All the applications we mentioned using 
SU(2) cluster coordinates can be extended to higher rank case and we plan to study them in the future.

The paper is organized as follows. In section II, we give a light review of the  theories of class ${\cal S}$ and put emphasis on the 
S duality behavior since such property is very important for constructing the cluster coordinates.
In section III, a review of the cluster algebra structure of SU(2) theory is given and we introduce
an equivalent network based on the triangulation, this part is mainly served as the basis for the higher rank case. 
In section IV, the brane web construction for the three punctured theories are introduced which is then used
to construct a network on a triangle; The network on a triangle is used to find a quiver for three punctured and 
four punctured theory. Section V describes the generalization to sphere with arbitrary number of punctures and to the higher genus case. Finally,
we give a conclusion and point out various generalizations and applications of these cluster coordinates.

\section{Review of  theories of class ${\cal S}$}
Four dimensional  theories of class ${\cal S}$ are defined as the IR limit of six dimensional $A_{N-1}$ $(2,0)$ theory 
compactified on a genus $g$ Riemann surface with $n$ punctures $\Sigma_{g,n}$. These theories are superconformal field
theories and generically no conventional Lagrangian formulation can be written down.  However, a lot 
of properties about these theories can be understood from the geometry of  $\Sigma$ and 
the Hitchin equation defined on it.  The gauge coupling
constants are identified as the complex structure moduli of $\Sigma_{g,n}$, and the mass parameters
are encoded as the local data at the punctures which are classified by the Young Tableaux with total
number of  $N$ boxes from which flavor symmetry can be read. These two sets of parameters exhaust all the relevant deformations 
of the UV theory. The Seiberg-Witten curve characterizing the IR behavior on the Coulomb branch is 
identified as spectral curve of the Hitchin's fibration which is then determined by the complex structure constant and the local data on the puncture. 

If the 4d theories are further compactified on a circle with radius approaching to zero, the four dimensional theories
flow to a three dimensional ${\mathcal{N}}=4$ SCFT. These three dimensional SCFT has 
a mirror description: Star-shaped quiver gauge theories \cite{Benini:2010uu}. Such quiver gauge theories all have Lagrangian 
description and one can extract some information of four dimensional theory from it, for example,
the full flavor symmetry. The superconformal index of 
these theories are also calculated recently \cite{Gadde:2011uv}. 
 
\begin{figure}[htbp]
\small
\centering
\includegraphics[width=10cm]{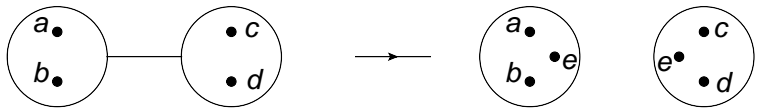}
\caption{A four punctured sphere is decomposed into two three punctured spheres in the degeneration limit.  A new puncture $e$ appears on
each three punctured sphere. }
\label{degeneration}
\end{figure}

The S duality property of these theories has a very nice geometric interpretation in terms of $\Sigma_{g,n}$: it is the modular group of the complex structure moduli space.
The weakly coupled gauge group description corresponds to the degeneration limit of the Riemann surface.  The weakly coupled gauge group
is regarded as living on the tiny long tube in the degeneration limit. 
The Riemann surface is decomposed into two parts $\Sigma_1$ and $\Sigma_2$ when the gauge coupling is completely turned off, and two identical new punctures $e$ appear
on $\Sigma_1$ and $\Sigma_2$, see figure.~\ref{degeneration}. 

\begin{figure}[htbp]
\small
\centering
\includegraphics[width=10cm]{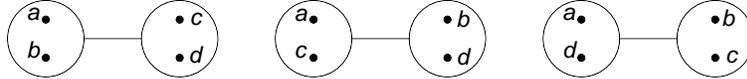}
\caption{Three duality frames for a SCFT defined by compactifying six dimensional $(2,0)$ theory on a  four punctured sphere. }
\label{duality}
\end{figure}

Physically, the original gauge theory can be thought of  as constructed by gauging the 
diagonal flavor symmetry of these two new punctures. In the completely 
degeneration limit, the punctured Riemann surface is decomposed into a  collection  of three punctured 
spheres. Generically, such three punctured sphere represents isolated strongly coupled SCFT which
plays the role of "matter" in constructing the full theory.   Different degeneration limits correspond to different 
duality frames in which the weakly coupled gauge groups and matter content are usually quite different.
For example, 
there are three weakly coupled duality frames which are in correspondence with three different 
degeneration limits for a four punctured theory. Each duality frame is regarded as colliding two punctures, see figure.~\ref{duality}.

The algorithm of determining the new punctures in the degeneration limit and the decoupled gauge group
is developed in \cite{Nanopoulos:2010ga}. The idea is to compare the Coulomb branch dimension of the original full theory and 
the decomposed theories after the degeneration.  Since this is important for our later development, let's
give a more detailed review below. 

The punctures are classified by a Young Tableaux with columns $[n_1,n_2,..n_r]$ which encodes $(r-1)$ mass 
parameters and the flavor symmetry group. One can define a N dimensional vector with entries
\begin{equation}
p_i=i-s_i,
\end{equation}
where $s_i$ is the height of $i$th box in the Young Tableaux.  It is easy to see that there is a one-to-one correspondence
between this vector and the Young Tableaux. We call a puncture "full" if the Young Tableaux is $[1,1,1..1]$ and 
 "simple" if the partition is $[N-1,1]$. 

Consider first the theory defined by putting an arbitrary number of punctures on a Riemann sphere, 
the total number of dimension $i$ Coulomb branch parameters is given by
\begin{equation}
d_i=\sum_jp_i^{(j)}-2i+1,
\end{equation}
where we sum over all the $i$th component of the vector.
We call a  theory "reducible" if there is certain integer $j$ such that $d_j<0$, otherwise it is called 
"irreducible" theory. If a "irreducible" theory has $d_N=0$, it is always possible to find a representation using $SU(M) (M<N)$ Hitchin
equation.

The new puncture in the degeneration limit is derived by comparing the Coulomb branch dimensions for the 
original theory and the decomposed theory. For example,  
there are two three punctured spheres and a decoupled gauge group in the degeneration limit of a four punctured theory. By matching the 
total Coulomb branch dimensions of these three pieces with the original theory, one can find the new punctures. 
An algorithm is developed after dealing with the subtle points associated with the "reducible" three punctured theory.

We give the result here and refer the interested reader to \cite{Nanopoulos:2010ga} for more details, see also \cite{Chacaltana:2010ks}.
Let's assume puncture $a$ and $b$ are colliding in the degeneration limit and define $\delta_{1i}=p_{i}^{(a)}+p_{i}^{(b)}$, 
 $\delta_{2i}=p_{i}^{(c)}+p_{i}^{(d)}$, the new puncture $e$ has the following vector
\begin{equation}
p_{i}^{(e)}=min(\delta_{1i},\delta_{2i},i-1).
\label{basic}
\end{equation}

The puncture $e$ is always the full puncture if both three punctured theories are "irreducible". If one of the three 
punctured theory say $abe$ is "reducible", then puncture $e$ is completely determined by the external two punctures $a$ and $b$.
The decoupled gauge group has a dimension $i$ operator if $\delta_{1i}\geq i$ in this case.  

The result is the same for the higher genus theory if the degeneration does not reduce the genus. If the genus is reduced in the degeneration limit, the 
new puncture is always the full puncture and the decoupled gauge group is always SU(N).

\section{Cluster coordinates, triangulation and SU(2) theory}

The cluster coordinates are a collection of triples $(\epsilon_{ij}, X_i, Y_i)$ where $\epsilon_{ij}$ is a skew-symmetrizable matrix and 
$x_i$ and $y_i$ are two sets of coordinates. Each triple is called a seed and two seeds are related by mutations. The antisymmetric
matrix $\epsilon_{ij}$ (we only consider this special case in this paper) can be represented nicely in terms of  a quiver:  there are $n$ total of quiver nodes and
the number of arrows from $i$th node to $j$th node is $\epsilon_{ij}$.

\begin{figure}[htbp]
\small
\centering
\includegraphics[width=10cm]{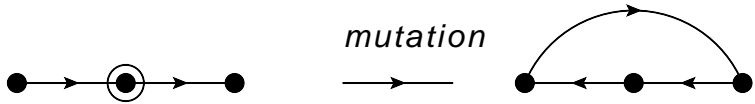}
\caption{The quiver mutation. }
\label{seiberg}
\end{figure}

Two different triples are related by the mutation on a given quiver node. The mutation formula for the matrix $\epsilon_{ij}$ is 
\begin{equation}
\epsilon^{'}_{ij}=\left\{
\begin{array}{c l}
    -\epsilon_{ij}& if~i=k~or~j=k\\
  \epsilon_{ij}+sgn(\epsilon_{ik})[\epsilon_{ik}\epsilon_{kj}]_+ & otherwise
\end{array}\right.
\end{equation}
where $[x]_+=max(x,0)$.  Basically, this formula means all the quiver arrows attached on the node $k$ is reversed and for any 
pair of quiver nodes $(ij)$, if there are $r$ oriented arrows from node $i$ to node  $k$ and $s$ oriented arrows from  node $k$ to node $j$, then 
after mutation,  a  total of $rs$ arrows between node $i$ and node $j$ is generated and the new quiver arrows are $(rs+\epsilon_{ij})$.
This formula can be most easily seen from the quiver, 
see figure.~\ref{seiberg}, it is interesting to see the quiver mutation has the same form as the Seiberg duality.

The transformation formula under mutation for coordinates $x_i$ and $y_i$ can be written appropriately. Since our focus in this paper is on the quiver mutation, we 
will not reproduce these formulas. The interested reader can find these formulas in \cite{Fomin2001}.

The cluster algebra structure of $\mathcal{N}=2$ SU(2)  theories are extensively studied in the literature. Here we will give a 
short review on how the quiver and quiver mutation can be nicely encoded into the triangulation of the punctured Riemann surface. Moreover,
a bipartite network is going to be introduced which is an equivalent to the triangulation. Such network is useful for our later 
study of the higher rank theory. 

We are interested in  a Riemann surface $\Sigma$ with genus $g$ and $n$ marked points (one can also consider Riemann surface with
boundary, this type of Riemann surface is relevant for Argyres-Douglas theory and asymptotical free theory; We will not study
these cases in this paper and leave it for the future).  The corresponding theory is a generalized SU(2) quiver gauge theory and 
formed by gauging the flavor symmetry of tri-fundamentals represented by a three punctured sphere. The total number of SU(2) gauge groups and the dimensions of Coulomb branch are 
\begin{equation}
n_r=n-3(1-g).
\end{equation}
Each puncture carries a SU(2) flavor group and there are a total of $n_f=n$ mass parameters.

The cluster coordinates are defined using the triangulation (see \cite{fomin-2008-201} for extensive description and the references there) which is achieved by using arcs on punctured Riemann surface.
A simple arc $\gamma$ is a curve on $\Sigma$ such that the end points of
$\gamma$ are the marked points and it does not intersect itself.  Two simple arcs are called compatible if they do not intersect in the 
interior of the Riemann surface.

\begin{figure}[htbp]
\small
\centering
\includegraphics[width=6cm]{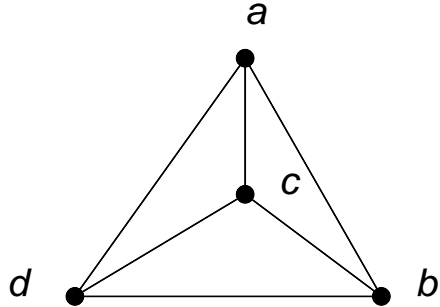}
\caption{A triangulation of four punctured sphere.}
\label{4sphere}
\end{figure}

An ideal triangulation is formed by picking a maximal collection of distinct pairwise compatible arcs. A triangulation for four punctured 
sphere is depicted in figure.~\ref{4sphere}. It is easy to see that the total number of edges for a triangulation are
\begin{equation}
n=6g+3n-6.
\end{equation}
This number is equal to $2n_r+n_f$, where $n_r$ is the Coulomb branch dimension and $n_f$ is the number of mass parameters.

\begin{figure}[htbp]
\small
\centering
\includegraphics[width=10cm]{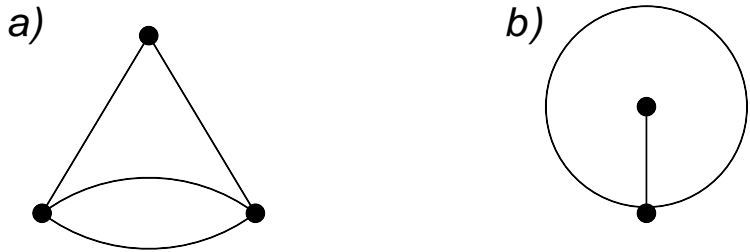}
\caption{Two special pieces in an ideal triangulation.}
\label{special}
\end{figure}

There are two special pieces for a triangulation as depicted in figure.~\ref{special}. The b type special piece is called self-folded ideal triangle, we will confine
ourselves to the triangulation without this piece.  A triangulation is called regular if there are no special pieces, we do allow type a piece though.

Now one can define a quiver or a anti-symmetric matrix from each triangulation. The matrix is found from the signed adjacent matrix of the edges:
first, assign one quiver node for each edge; second, there
is an arrow connecting two nodes if these two edges are in the same triangle.The orientation of the arrows are determined by taking a orientation of the triangle, i.e. clockwise direction. If 
the edge $i$ is before the edge $j$ in the same triangle, then $\epsilon_{ij}=1$. The total number of the arrows between two quiver nodes are the signed sum if the corresponding edges
are in more than one triangle.

\begin{figure}[htbp]
\small
\centering
\includegraphics[width=10cm]{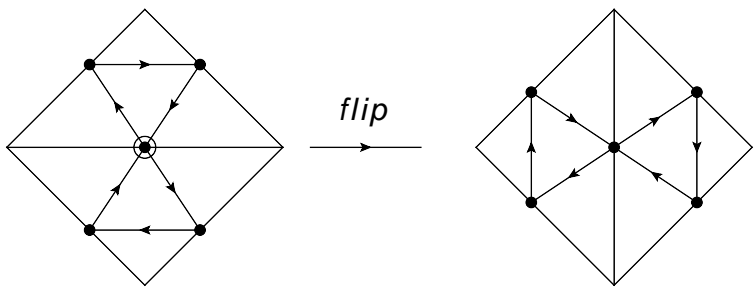}
\caption{Flip on a triangulation of the quadrilateral, the corresponding quiver is also drawn. The quiver on the right is derived by mutating the circled node of the left quiver.}
\label{flip}
\end{figure}

The ideal triangulation is not unique and different triangulations are related by a composition of  local move: flip.  The flip relates two triangulations of the quadrilateral, see
figure.~\ref{flip}.  One can check the two quivers for the two triangulations are related by the quiver mutation on the node representing the diagonal edge.

Although we do not study the cluster $x$ and $y$ coordinates in this paper, we should point out that these coordinates both have the geometric meaning. The x coordinates are the so-called Penner coordinates and the $y$ coordinates are the so-called Fock-Goncharov (or Thurston) coordinates. 

\begin{figure}[htbp]
\small
\centering
\includegraphics[width=10cm]{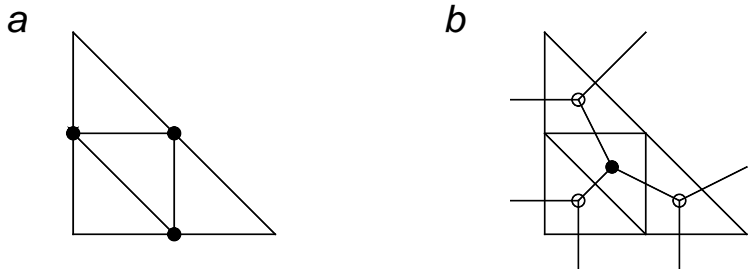}
\caption{A bipartite network structure on a triangle of the triangulation.}
\label{Su(2)}
\end{figure}

Now we want to introduce another structure  on Riemann surface using the triangulation. Let's put one more dot on each boundary of triangle and triangulate the big triangle as described in figure.~\ref{Su(2)}.  Let's put a black
vertex on the middle small triangle and white vertex on other triangles. The network is formed by the following rules: connecting white vertex and black vertex if there are common edges for the two small triangles containing the vertices. 
Finally,  there is a line coming out of the boundary of the big triangle for the vertex which has an edge of boundary.

\begin{figure}[htbp]
\small
\centering
\includegraphics[width=8cm]{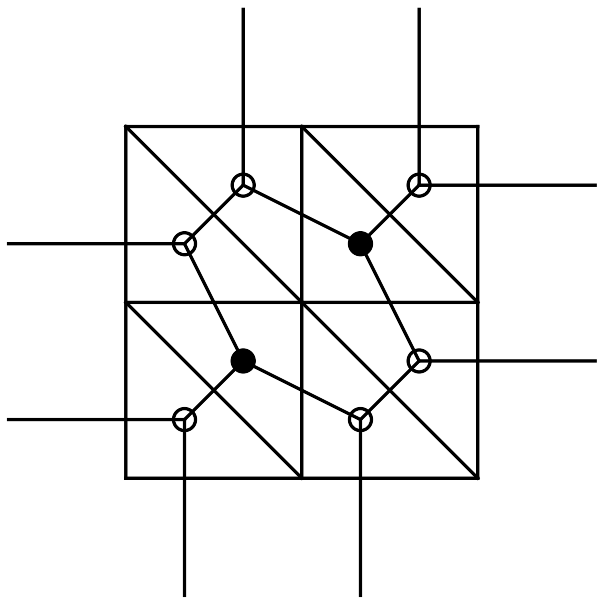}
\caption{Gluing networks  of two triangles to form the network of a quadrilateral.}
\label{network1}
\end{figure}

The full network is formed by gluing network of each triangle. The gluing rule is very simple: if two triangles share a common edge, we simply connect the boundary lines
of the network on each triangle, see figure.~\ref{network1}. Now we can forget the triangulation and discuss everything in terms of the network. Such type of network has
been studied in dimer and $\mathcal{N}=1$ gauge theories context. One can also read a quiver from the network: Each surface represents a quiver node, and each 
black dot gives an oriented  cycles of the arrows between the surfaces sharing this node. There is no quiver node for the surface circling around the punctures though.

\begin{figure}[htbp]
\small
\centering
\includegraphics[width=10cm]{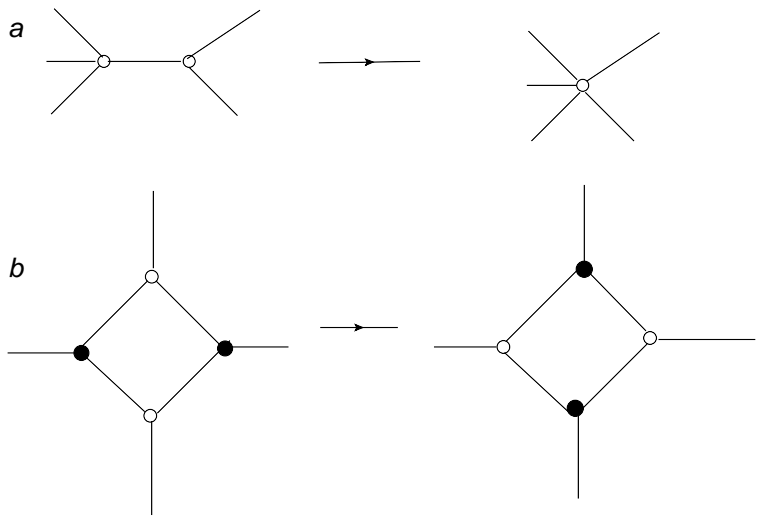}
\caption{a): Contracting the edge connecting vertices with same color; b) Square move.}
\label{square}
\end{figure}

As discussed earlier, the triangulation is not unique and different triangulations generate different networks. Since the triangulations are related using flips, we would like
to have a representation of flip in terms of network transformation. Such local transformation  shown in figure.~\ref{square}b  is called square move (or spider move) in the literature.

To do that,  one need to first define a trivial (un)contract operations to make the glued network bipartite: if there is an edge connecting two vertices with the same color,
one can shrink this edge and merge these two vertices together. The contracting operations can be done on two white dots on the diagonal edge
 as described in figure.~\ref{square}a, the surface on the internal edge of figure.~\ref{network1} is becoming a square and
a square move can be done. It is easy to check the transformed network is the same as the network defined on the triangulation after the flip.

So the cluster coordinates can be read either from a triangulation or a network on the punctured Riemann surface. The quiver mutation is 
represented geometrically as the flip in the triangulation case and the square move in the network case.

\section{Cluster coordinates for higher rank theory}
\begin{figure}[htbp]
\small
\centering
\includegraphics[width=4cm]{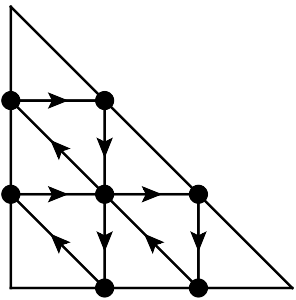}
\caption{The quiver on a triangle of SU(3) theory with full punctures.}
\label{Su(3)}
\end{figure}

The cluster coordinates for SU(2) group are nicely defined using the triangulation of the punctured Riemann surface and
such constructions are well developed in studying Teichmuller theory.
A remarkable generalization to higher rank SU(N) theory with all full punctures are given by Fock and Goncharov (FG). 
They also start with a triangulation of punctured Riemann surface but put more structure in each of the triangles. What they are 
doing is basically the generalization of what is depicted in figure.~\ref{Su(2)} in which the triangle is further triangulated by putting
one more point on each boundary.  Here let's put $(N-1)$ points on each boundary and 
do the triangulations in a very natural way, see figure.~\ref{Su(3)} for an example. Now associate a quiver node for each marked points
on the boundary or inside the triangle, and the quiver arrows are determined by the little triangle with opposite orientation of 
the original big triangle.  The whole quiver is derived by naively identifying the boundary nodes of two glued triangles.  Different 
triangulations are related by the local flip move which is represented by a sequence of quiver mutations on the quivers.  We will leave
more details later and just point out that there is a very parallel story with SU(2) theory.

However, unlike SU(2) which has only one type of punctures, there are many other types of punctures besides the full puncture for the 
higher rank theory. No such cluster coordinates construction has been given in the literature which plays an 
important role in various physical applications, i.e the Argyres-Seiberg duality.  In the following, a combinatorial method is developed to 
find cluster coordinates for non-full punctures.

Naively, the first step for constructing cluster coordinates for generic puncture should be similar as the full puncture case: starting with
the triangulation and do further tessellation on each triangle. There is actually already a type of tessellation coming from the study of 
 theories of class ${\cal S}$ using brane construction \cite{Benini:2009gi}.  The first hint that this type of tessellation might be 
the correct one is that it is exactly the one used by Fock and Goncharov for the full puncture case.  Below we will 
use this tessellation to construct the cluster coordinates for the non-full puncture and show many convincing evidence 
that our construction is a generalization of Fock and Goncharov's result.

\subsection{Brane web, network and quiver}

\subsubsection{Irreducible theory}
It was noted in \cite{Benini:2009gi} that the five dimensional version of three punctured theory can be engineered using brane webs of Type IIB string theory.  
Such brane constructions provide a very nice geometric picture which is used to probe properties of the corresponding SCFT,  for example,
 the  Coulomb branch and Higgs branch dimensions can be calculated easily.  Since the 4d theories are in general strongly coupled and usually has no lagrangian description, 
such geometric construction turns out to be a quite useful tool. 
We briefly review their constructions below and clarify some points which is not explicitly discussed in that paper. 
In particular, a detailed study of reducible theory is given and a geometric origin for formula [\ref{basic}] is found.

\begin{figure}[htbp]
\small
\centering
\includegraphics[width=4cm]{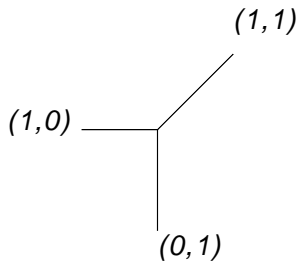}
\caption{The basic building block for the brane web. }
\label{block}
\end{figure}

The brane web is an intersecting brane configuration of D5 brane, NS5 brane and $(1,1)$ five brane of Type IIB string theory. One
also need to put D7 branes on which multiple five branes can be terminated to produce the generic puncture.  For the generic puncture,
more than one five branes are terminated on a single D7 brane while  only one five brane is ending on a given D7 brane for the full puncture.
The pattern of D5-D7 intersection is completely determined by the corresponding  Young
Tableaux of punctures. The basic building block of the brane web is the trivalent vertex (brane junction) connecting three types of five branes, as shown in figure.~\ref{block}. 

The brane web for SCFT formed by three punctures can be most efficiently constructed using a dot diagram which is constructed using the Young Tableaux.
Concretely, let's first fix a two dimensional integer lattice generated by two vectors  $(0,1)$ and $(1,0)$, then
draw a triangle with vertices $(0,0), (0,N), (N,0)$ which represents three punctures and labeled as $(a,b,c)$.
There are essentially two inequivalent ways of labeling the vertices which correspond
to two cyclic order of three letters: $a\rightarrow b \rightarrow c$ and $b\rightarrow a \rightarrow c$. We just choose one of them. 

\begin{figure}[t]
\small
\centering
\includegraphics[width=10cm]{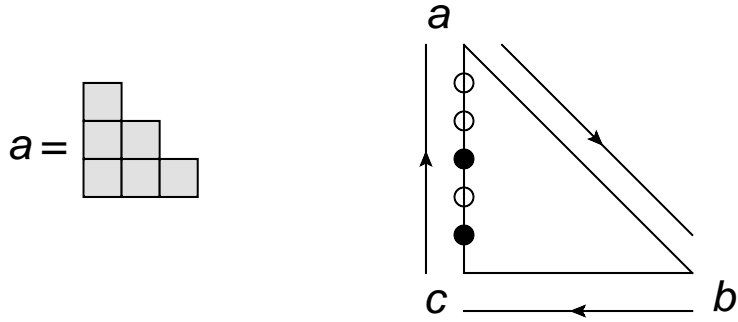}
\caption{This figure shows how the information of punctures are projected on the boundary edges. The small segments
formed by black dots has the same lengths as the height of the column of Young Tableaux.   }
\label{dot}
\end{figure}

The lattice points bounded by the triangle (including the points on the boundary) will be colored as either black or
white. The vertices of  triangle are always colored as black and
the color pattern on the boundary edges  are determined by the Young Tableaux using the following rule. 
Our convention is to project  the information of a puncture to the edge right before it, i.e.
the color pattern on edge $ac$ is dictated by puncture $a$. 
if  $a$ has Young Tableaux with columns $[n_1,..n_r]$, then the black points are 
arranged such that the edges connected by black points has the length $n_r, n_{r-1}, ...n_1$ starting from vertex $c$.  Here only
black dots are regarded as the end points of the small edges, see figure.~\ref{dot} for an example.

The colors for the lattice points inside the triangle are determined by the preservation of supersymmetry: the so-called
s-rule. Basically, the s-rule says that some brane configurations are non-supersymmetric and discovered originally in the D3-D5-NS5
brane configurations.  s-rule states  that the configuration with more than one D3 branes ending on a given NS5 brane and D5 brane pair
is non-supersymmetric. The s-rule in the present context  means that only one five brane is allowed to end on a given seven brane
and five brane pair.  For more details on the discussion of the s-rule for the brane web, see \cite{Benini:2009gi}. We  summarize
their results and  show how to practically use the rules to construct a dot diagram from the information of the punctures.

The s-rule is used to decorate the internal points and the color pattern is used to tessellate the triangle using minimal polygon: connecting
the adjacent black dots to form the edges of the polygon. The edges of the polygon should be parallel with the boundary edges.
The s-rule forces the polygon to be the following two types:  triangle and trapezium and they have to satisfy the following
two extra constraints:

1. The three edges of a triangle have the same length $n$.

2. The four edges of the trapezium should have the length $n_2, n_1, n_2-n_1, n_1$, and the edges with
length $n_2$ and $n_2-n_1$ should be parallel.

It is important to know whether a consistent tessellation exists given three generic punctures.
We argue that there is always a consistent tessellation for irreducible theory.  Let's assume that the three punctures have Young Tableaux $a=[n_1,...n_r]$, 
$b=[l_1,....l_s]$ and $c=[m_1,...m_t]$.  The top entry  $p_N$  of the vector derived from a puncture  is 
$p_N=N-h_1$, where $h_1$ is the height of the first column.  Remember the "irreducible" condition implies that there is no negative naive 
Coulomb branch dimensions calculated from three punctures.  Imposing this condition on dimension $N$ operators, we have
\begin{equation}
\sum_{i=1}^3p_N^{(i)}-2N+1\geq 0\rightarrow  n_1+l_1+m_1\leq (N+1).
\label{constraint}
\end{equation}
This means that the sum of any two top heights are no more than $N$, say $n_1+l_1\leq N$.

\begin{figure}[htbp]
\small
\centering
\includegraphics[width=10cm]{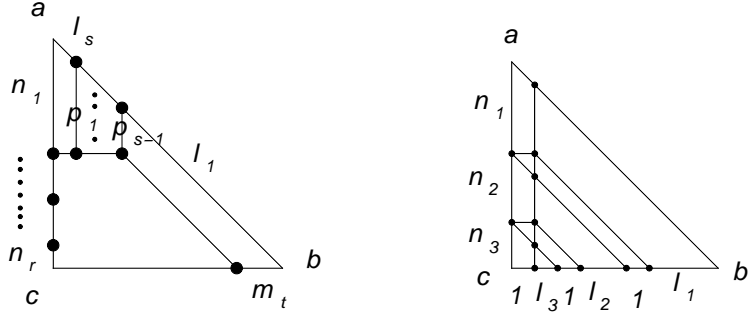}
\caption{Left: a tessellation of a triangle using the information of Young Tableaux. To satisfy our constraint, 
we must have $m_t=p_{s-1}$. Right: A consistent tessellation with $\textbf{b}$ a simple 
puncture and $\textbf{a}$ a generic puncture with three columns. $l_1=n_1-1$, $l_2=n_2-1$
and $l_3=n_3-1$.}
\label{dotpattern}
\end{figure}

Let's first decorate the colors for the boundary points using the information of the puncture. Now assume without loss of generality that $n_1\geq l_1\geq m_1$ 
and start do the tessellation starting from the vertex $a$ with small edges $n_1$ and
$l_s$ connecting it. Part of tessellation is depicted in the left of figure.~\ref{dotpattern}. Let's further assume $n_1>l_s$ (the boundary situation
$n_1=l_1=l_s$ can be treated with some care, our conclusion is still valid though), then
 the first trapezium must have another line $p_1$ parallel with $ac$ from the constraint which has length $p_1=n_1-l_s$. 
 In the second step, if $l_{s-1}<p_1$, then $p_2$ is still parallel with $ac$, and has lengths $p_2=n_1-l_s-l_{s-1}$, etc.  

 We can continue doing tessellation along edge $ab$ without any difficulty.
 The trouble appears if the tessellation has the structure shown in the left of figure.~\ref{dotpattern}.  Such situation can appear if
 $p_{s-2}>l_2$, which  means that $p_{s-1}$ is also in parallel  with edge $ac$.  The condition on $p_{s-2}$ implies:
\begin{equation}
p_{s-2}=n_1-(l_s+...l_3)>l_2 ~~\rightarrow N-l_1< n_1.
\label{constraint1}
\end{equation}
In the second step, the condition $\sum_j l_j=N$ is used. If the above situation
happens, $m_t$ has to satisfy the following condition
\begin{equation}
m_t=p_{s-1}=n_1-(l_s+l_{s-1}+...+l_2)=n_1+l_1-N.
\label{consistent}
\end{equation}
So in this situation, the third puncture is not arbitrary and the last column 
$m_t$ is completely fixed to have a consistent tessellation! Similar reasoning shows that actually the third 
puncture is completely fixed in the above situation. Fortunately,  $n_1+l_1\leq N$ is guaranteed for irreducible theory  as  can be seen from formula (\ref{constraint})
and the above troublesome situation will never happen.

Similar analysis can be applied to other two corners and no contradiction  to s-rule
is found if the theory is irreducible. Once  a consistent tessellation along 
the boundary edge can be found, it is easy to see that a full consistent tessellation can also be 
achieved. Our conclusion is that there is always a consistent tessellation for irreducible theory.

\subsubsection{Reducible theory}
However, we do need to consider reducible theories since  they 
are inevitable in understanding $S$ duality of gauge theory.  The reducible condition means 
that the internal puncture in the degeneration limit is completely determined by the data on
the two small colliding punctures.  Let's assume these two small colliding puncture has the partition 
$a=[n_1,....n_r]$ and $b=[l_1,....l_s]$, and apply the "reducible" condition to the dimension $N$ operators:
\begin{equation}
\delta_{1N}=p_N^{(a)}+p_N^{(b)}\leq N-1\rightarrow n_1+l_1 \geq N+1.
\end{equation}
Interestingly, in such a situation, the third puncture is completely fixed if we 
want to have a consistent tessellation as seen from formula (\ref{consistent}).
This is not a coincidence and it appears the third puncture in our current
graphic approach is the same as derived from formula (\ref{basic}).

It may be helpful to  look  at a simple example shown on the right of figure.~\ref{dotpattern}. The puncture $b$ is a simple puncture and puncture
$a$ has partitions $[n_1,n_2,n_3]$. From our formula $(\ref{basic})$,  the combining 
puncture has Young Tableaux $[n_1-1, n_2-1,n_3-1,1,1,1]$ (it is  formed by combining the first
two rows of the generic puncture into a single row). On the other hand, the only consistent tessellation with above two punctures is 
drawn on the right side of figure.~\ref{dotpattern} with 
\begin{equation}
l_1=n_1-1,~~l_2=n_2-1,~~l_3=n_3-1. 
\end{equation}
So the result from the consistent tessellation is equivalent to the result  by counting Coulomb branch dimensions.  Notice that  
lengths  pattern of the small edges on the segment $bc$ is not increasing monotonically as the other two
boundary edges.  This is not a problem since it is a internal puncture.

Generally speaking, the third puncture is completely fixed by the data on two "small" punctures 
to have a consistent tessellation. The  result using dot diagram is 
the same as that found using the  degeneration limit of Riemann surface as it should be.

\subsubsection{Brane web from dot diagram}
Once a consistent tessellation of the dot diagram has been found, a brane
web diagram can be constructed by mapping lines to orthogonal 5-branes. Area 1/2 triangles are mapped to the
usual junction of three 5-branes. The other minimal polygons are mapped to intersections
of 5-branes in which, because of the s-rule, a 5-brane cannot terminate on another one and
has to cross it. We say that the 5-brane jumps over the other one, even though there is no
real displacement.

Let's illustrate these brane constructions using some simple examples.  We 
use SU(3) theory examples and the general case is similar.
There are only two types of punctures in SU(3) theory: the minimal puncture with Young Tableaux
$[2,1]$ and the full puncture with Young Tableaux $[1,1,1]$.

\begin{figure}[htbp]
\small
\centering
\includegraphics[width=10cm]{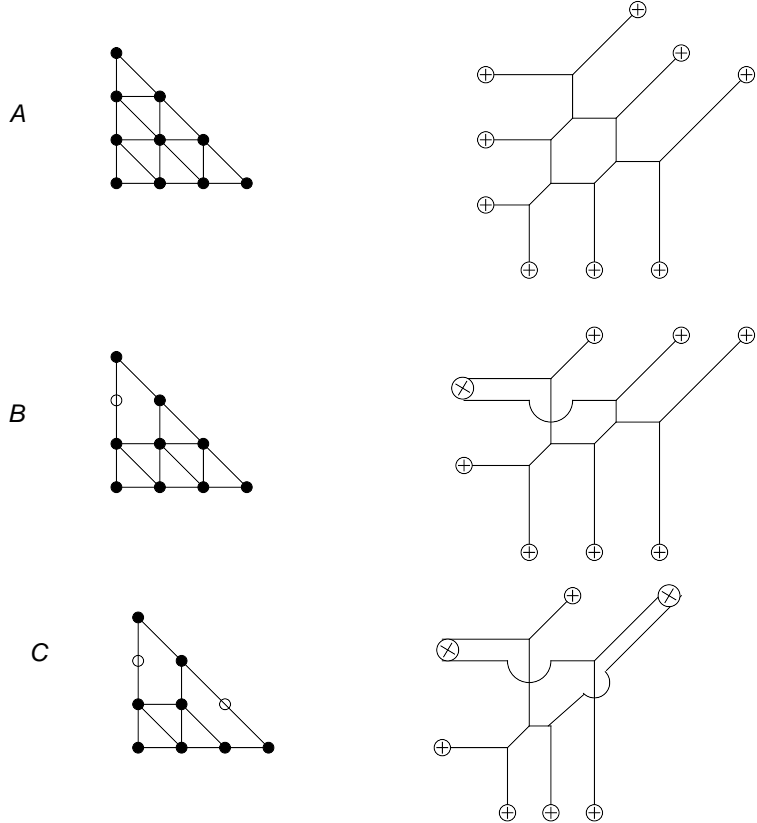}
\caption{A: The dot diagram and brane web diagram for sphere with three full puncture of SU(3), there is
one closed surface in brane web diagram so the Coulomb branch dimension is one. B: The dot diagram and web diagram for
the bi-fundamental, there is no closed surface so this theory is a free theory. C: The dot and web diagram for 
a theory with two simple punctures and one full puncture.
}
\label{E6}
\end{figure}

\textbf{\emph{Example 1}}: The three punctures are all of full type, which actually represents $E_6$ theory which is an isolated SCFT with $E_6$ flavor symmetry. The dot
diagram and the web diagram are shown in figure.~\ref{E6}A. The dot diagram is  the one used by Fock-Goncharov in constructing the cluster coordinates for SU(3) group.
The closed surface in the web diagram represents the Coulomb branch parameter. There is only one
closed surface  representing one dimension Coulomb branch which is in agreement with the field theory
result.

\textbf{\emph{Example 2}}: Two full punctures and a simple puncture are used to represent
the bi-fundamental hypermultiplets of SU(3) group. The dot diagram and brane web are shown in figure.~\ref{E6}B. There is no closed
surface in the web diagram, so the Coulomb branch is zero and it is a free field theory. It is interesting to note that
the two full punctures in the brane web have different configurations. This seemingly strange fact is in fact
very important for us as we will show later.

\textbf{\emph{Example 3}}: This theory consists of a Riemann sphere with two simple punctures and one full puncture. 
This is an important ingredient in understanding Argyres-Seiberg duality.
The dot diagram and the web diagram is shown in figure.~\ref{E6}C. 

There is no consistent tessellation for the theory with three simple punctures from our earlier analysis on reducible 
theory so that above three examples are all  what we need for SU(3) theory.

\newpage 
\subsubsection{Network and quiver from brane web}
\begin{figure}[htbp]
\small
\centering
\includegraphics[width=10cm]{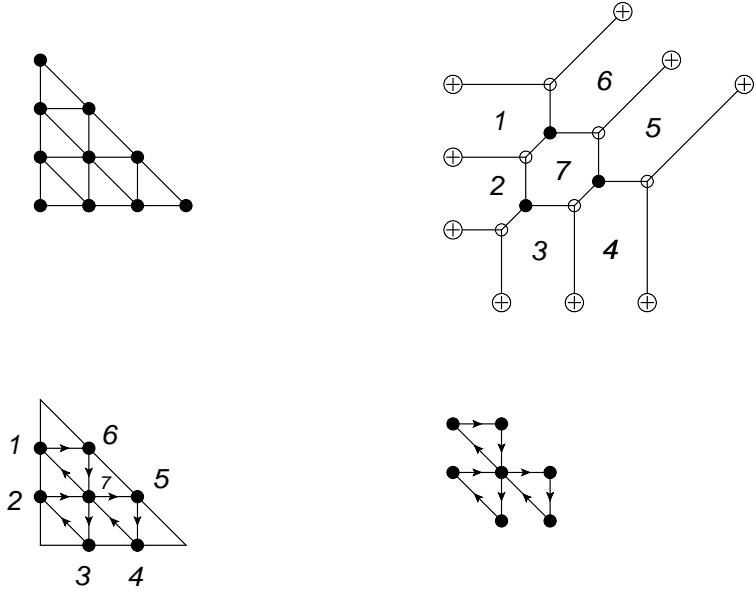}
\caption{The brane web of $E_6$ theory which can be regarded as a bipartite network. The quiver is read straightforwardly from the network.
}
\label{E6package}
\end{figure}

The brane web diagram is exactly the network for the full puncture cases if we assign appropriate color to the  brane junctions. 
  The color
 of the vertices is assigned using the following rule: a white color is put on the brane junction if the small triangle is in the same orientation of the big triangle, 
  otherwise we put a black color for the junction. The web diagram becomes a bipartite network and a quiver can read easily.
  The example of  $E_6$ theory is shown in figure.~\ref{E6package}.  

The generalization to the non-full puncture case is not good at first sight since the brane web
has one unpleasant feature: there are lines crossing each other. So the brane web itself
is not a network and simplification is needed.

\begin{figure}[htbp]
\small
\centering
\includegraphics[width=10cm]{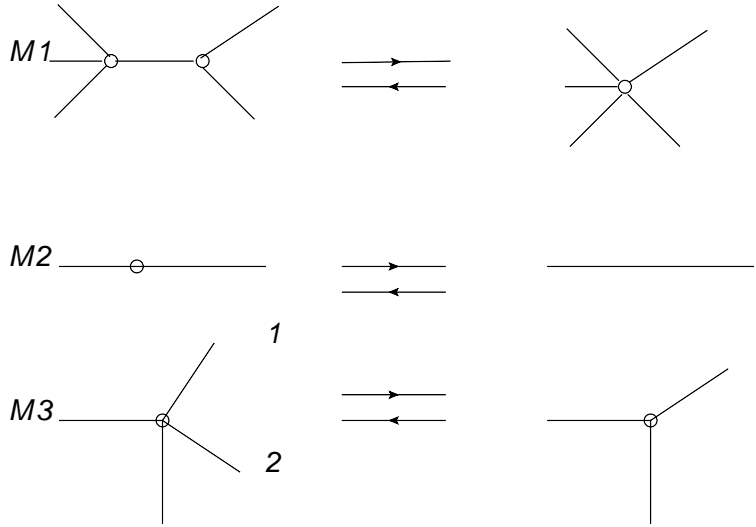}
\caption{M1 move contracts the edges connecting vertices with same color. M2 move remove the vertex with only 
two edges on it. M3 move remove one of the external edges ending on the same vertex, these external edges have to 
be coming out of the same boundary of the big triangle.
}
\label{Move1}
\end{figure}

To do the simplification, let's recall that the non-full puncture can be regarded as "higgsing" the full
puncture. At the level of network, the effect of "higgsing" is simply to remove one of the edge of the original 
network.  The network is not bipartite anymore and the following three moves depicted in figure.~\ref{Move1} can be 
 used to simplify the network and turn it into a bipartite network. M1 has already been discussed in earlier section which is used to 
contract edges connecting two vertices with same color. M2 move states that a vertex can be removed if it has only two edges attached on it.
 M3 move means that  one of the edges if the edges $1$ and $2$  can be removed if they are coming out of the same boundary of 
big triangle and attached on the same vertex. Notice that, M3 move only applies
to the edges coming out of a single boundary and is not allowed if $1$ and $2$ are belonging to separate 
boundaries. 

\begin{figure}[htbp]
\small
\centering
\includegraphics[width=10cm]{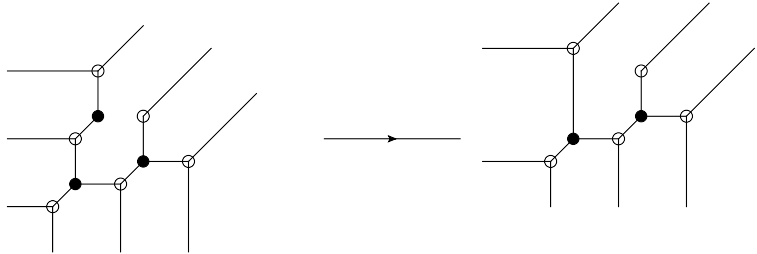}
\caption{Left: Remove one of the edges from the closed surface of a network. Right: the bipartite network after first removing the black vertex with
only two edges, then combining two white vertices and finally removing one of the boundary edges ending on the same vertex.
}
\label{edgerem}
\end{figure}

Let's consider the network of $E_6$ theory and remove one of the edges on the closed hexagon, as shown in
left of figure.~\ref{edgerem}.  One of the black dots can be removed using M2 move since there are only two edges ending on it.
Then use M1 move to contract the edges connecting two white vertices and M3 move to eliminate one of the boundary edges
ending on the same vertex, we are left with the network shown on the right of figure.~\ref{edgerem}. 
The  network on the right of figure.~\ref{edgerem} is almost like the brane web configuration with two full punctures and one simple 
puncture of SU(3) shown in figure.~\ref{E6}B. The only difference is that the brane leg crossing with other edges is removed. This motivates us to 
propose that a network without crossing can be derived from brane web by simply ignoring the brane legs  with crossing to others. In fact, 
the network is much easier to draw than the corresponding brane web. 

\begin{figure}[htbp]
\small
\centering
\includegraphics[width=10cm]{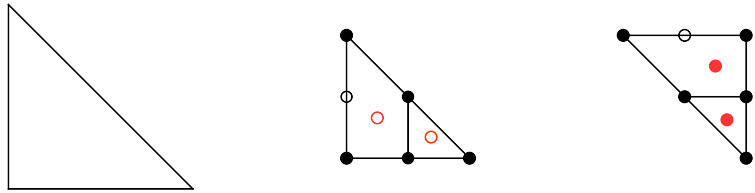}
\caption{Left: The orientation of the big triangle.  Middle: Put a white vertex to each polygon whose triangle completion has  the same orientation as 
the big triangle. Right: Put a black vertex to each polygon whose triangle completion has opposite orientation.}
\label{vertex}
\end{figure}

To state our rules for constructing the bipartite network, we need to distinguish two types of polygons in the tessellation of the big triangle:
The type A polygon is the one whose triangle completion has the same orientation as the big triangle, and
the triangle completion of the type B polygon has opposite orientation as the big triangle.  The colored vertices in each polygon of the tessellation are determined in the following way (see figure. \ref{vertex}):

a: Assign a white vertex to each type A polygon. 

b: Assign a black vertex to each type B polygon.
 
 A network is formed by connecting the white vertex and black vertex if there is a common edge between two corresponding polygons. 
 We never connect two vertices with same color even if the corresponding polygons have one common edge.  Finally, there is one line  coming 
 out of boundary for the vertex inside the polygon which has one piece of boundary edge of the triangle.  There are a total of $r$ outgoing edges for each boundary, where
 $r$ is number of columns in the corresponding Young Tableaux. The network formed in this way is always bipartite but there may be vertices with only two edges. 
 We can use various moves to remove these degree-2 vertices and  get another bipartite network from which a quiver without two cycles can be read in the following two steps:
 
 Step1: Remove degree two vertices and then use the contraction to form another bipartite graph.
 
\begin{figure}[htbp]
\small
\centering
\includegraphics[width=6cm]{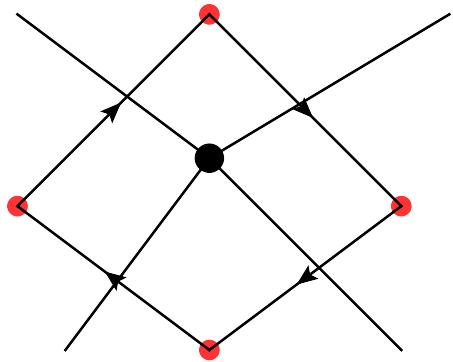}
\caption{The rule for finding the quiver arrows from the bipartite network.}
\label{quiverarrow}
\end{figure}

 Step 2:  Assign a quiver node to each surface and the quiver arrows are determined by the black vertices: there is 
 a clockwise closed circles around it, see figure. \ref{quiverarrow}.
  
Let's illustrate our construction of network using the following simple example.

\begin{figure}[htbp]
\small
\centering
\includegraphics[width=10cm]{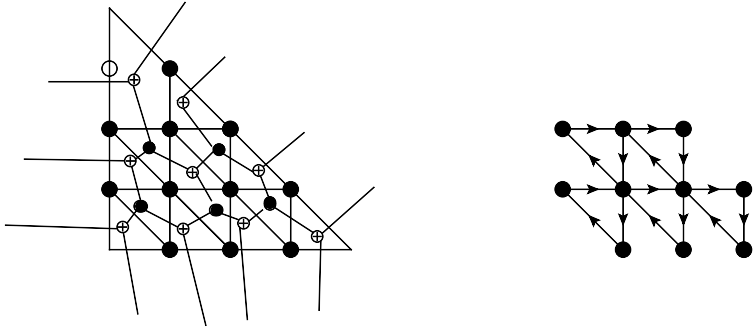}
\caption{Left: The dot diagram and separated regions for a theory with three punctures $Y_1=[1,1,1,1],~Y_2=[1,1,1,1],~Y_3=[2,1,1]$.
Right: the quiver from the dot diagram, associate one quiver node to each surface and the quiver arrows are read from the black vertices. }
\label{networkexample}
\end{figure}

\textbf{Example}:  The three punctures have partitions $Y_1=[1,1,1,1],~Y_2=[1,1,1,1],~Y_3=[2,1,1]$. The dot diagram and 
the network are depicted on left of figure.~\ref{networkexample}.  The quiver is also shown on the right of figure.~\ref{networkexample}.

\newpage

\subsubsection{Three punctures}
\begin{figure}[htbp]
\small
\centering
\includegraphics[width=6cm]{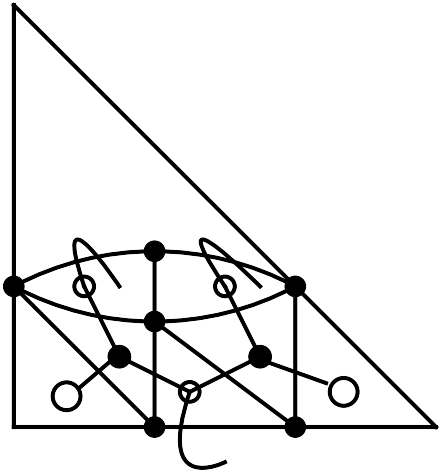}
\caption{Glue two networks for two triangles of SU(3) bi-fundamentals. }
\label{glue}
\end{figure}

 It is not hard to get the full network of a three punctured theory  by gluing the network of two triangles in the triangulation which are bounded by the same three edges.  
The gluing pattern around the puncture is best described by extending the network into a 
three dimensional sphere. For instance, consider the bi-fundamental hypermultiplets of SU(3) theory, there are two white dots around 
the special puncture and no lines connecting them. Let's  put these two white dots around the special puncture inside the three sphere as shown in figure.~\ref{glue}, and 
identify the white dots on both triangles. The full quiver is actually simple to get: the quiver of the triangle on the back is derived by reversing all the quiver arrows 
and the full quiver is derived by identifying the boundary nodes corresponding to the open surfaces of network of one triangle.

\begin{figure}[htbp]
\small
\centering
\includegraphics[width=10cm]{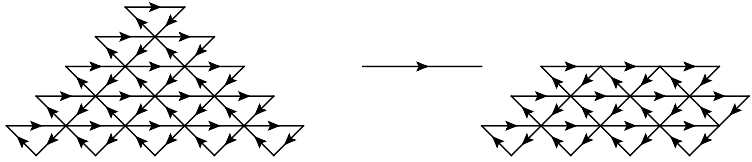}
\caption{The quiver for generic puncture can be derived from full puncture by simply eliminating first $(a_1-1)$ rows. Here $a$ is the puncture with partition $[3,1,1,1]$ and
$b$ has partition $[2,1,1,1,1]$.}
\label{ful_red}
\end{figure}

Our construction is based on a cyclic order of the punctures and there is another equally good choice. These two choices have to be isomorphic  to ensure the consistency. 
Now let's prove this statement starting from the simplest examples for which the three punctures have the partition $[n_1,1,1,1,..1]$, namely, only the height of first column
in the Young Tableaux is larger than one.  
The quiver of this special configuration is easy to get from the quiver of three full punctures. The quiver of three full punctures 
are symmetric for three punctures as shown on the left of figure.~\ref{ful_red}. If the height of  the first column of puncture $a$ is $a_1$, then the quiver is derived by 
simply eliminating the first $a_1-1$ rows. The quiver for $a_1=3$ and $b_1=2$ is shown on the right of figure.~\ref{ful_red}.  

It is easy to see for this special configurations of three punctures, the quiver for the cyclic order $abc$ is derived by  reversing all the quiver arrows of quiver from cyclic order $acb$.
The overall quiver  orientation  does not matter since there is always another quiver with opposite orientation on the back triangle. Therefore the full quivers are the same 
for the cyclic order $abc$ and $acb$ for this special case.

\begin{figure}[htbp]
\small
\centering
\includegraphics[width=8cm]{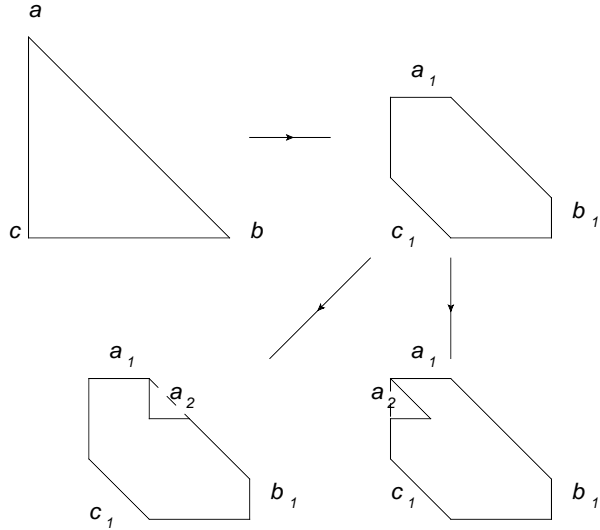}
\caption{Upper: The effect of a non-full puncture with only one non-unit column is to cut the triangle into a hexagon and the quiver is derived by filling small area 1/2 triangles into it. Lower: if puncture $a$ has partition $[a_1,a_2,1,1,1]$,
the effect is to cut another triangle from the original boundary of the triangle. There are two choices depending on which cyclic order of punctures are chosen. The final quivers are the same
if the new cut is not overlapping with the other cuts. }
\label{cut}
\end{figure}

The effect of the non-full puncture is to cut a $n_1\times n_1$ triangle from the big triangle from vertex representing the puncture as shown schematically in figure.~\ref{cut}. The quiver is derived by filling
small area ${1\over2}$ triangles into the remaining polygon. This picture does not depend on which cyclic order is chosen to define the quiver.  

Now let's further assume puncture $a$ has the partition $[a_1,a_2,1,1,...1]$, namely, the height of second column is also bigger than one.  Moreover, we assume 
$a_2\leq N-a_1-b_1$ and $a_2\leq N-a_1-c_1$ (the general case is similar). Similarly, the effect of the second column is to remove another $a_2\times a_2$ triangle on the edge of the 
hexagon which is part of the boundary of the original triangle. There are two choices which correspond to put the cut between puncture $a$ and puncture $b$ or
between puncture $b$ and $c$, which correspond to two different cyclic orders of the punctures. The quivers defined by two cuts are different and one possible 
reason is that they define the different decorated space although the underlying moduli space of the flat connection is the same, this however deserves 
further study.  The total number of quiver nodes are $2n_r+n_f$ from our construction and we conjecture they are the BPS quiver for the underlying
SCFT.

\newpage
\subsection{Quadrilateral and four punctures}
Things become more interesting for Riemann sphere with four punctures and we only consider theories with nonzero dimension $N$ Coulomb branch parameter in this subsection. 
There is an interesting S duality behavior which can be understood from the different degeneration limits of the Riemann surface. 

The natural guess for building a network is the following: start with a triangulation and construct a network for each triangle, finally glue them to form a full network.
However, there is an immediate difference from the three
punctured case: the number of edges in the triangulation are bigger than the number of punctures, so it is important to know what kind of information should be put on the internal edges.

The idea is coming from how the corresponding  theories of class ${\cal S}$ are built from three punctured
sphere. The punctured Riemann surface has a pants decomposition: it can be decomposed into
various three punctured sphere, the full gauge theory is derived by gauging the diagonal
flavor symmetry of the new appeared punctures.  The basic object in the triangulation is the quadrilateral which 
is built from gluing two triangles.  In analogy with the gauge theory construction, it is natural to decorate the boundary of 
quadrilateral using the information of four punctures, and decorate the internal edges using the information of 
the newly appearing puncture in the degeneration limit. 

In the gauge theory side, the new puncture is calculated by ensuring that the total number of Coulomb branch of various pieces after
complete degeneration is the same as that of the original theory. If we decorate the internal edge using the same puncture,
then naively the total number of quiver nodes are $2n_r+n_f$ which is exactly what we need.

As in the triangle case, one need to take a cyclic order of the four punctures. Let's assume that the the four punctures are arranged in
the cyclic order $(a,b,c,d)$.  The quadrilateral is decomposed into two triangles with vertices $(a,d,e)$ and $(b,c,e)$. The puncture $e$ is 
given by the gauge theory calculation, i.e. the duality frame in which puncture $a$ and $d$ are colliding each other. 
\begin{figure}[htbp]
\small
\centering
\includegraphics[width=8cm]{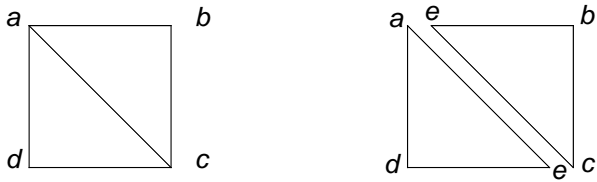}
\caption{A quadrilateral is decomposed into two triangles with two new vertices labeled by e.  }
\label{decomp}
\end{figure}

If the new puncture is non-full, then one of the decomposed three punctured theories is reducible and the decoration on the internal 
edge is completely fixed by the two punctures of the reducible theory and has the unusual ordering. The decoration on the glued edge of the other triangle
has the same ordering.

After determining the decoration on the internal edge, a sub-network can be constructed on each triangle which is used to
get the full network for the quadrilateral. A bipartite network can then be found from which an equivalent quiver can be easily created.

\subsubsection{Minimal network}
\begin{figure}[htbp]
\small
\centering
\includegraphics[width=10cm]{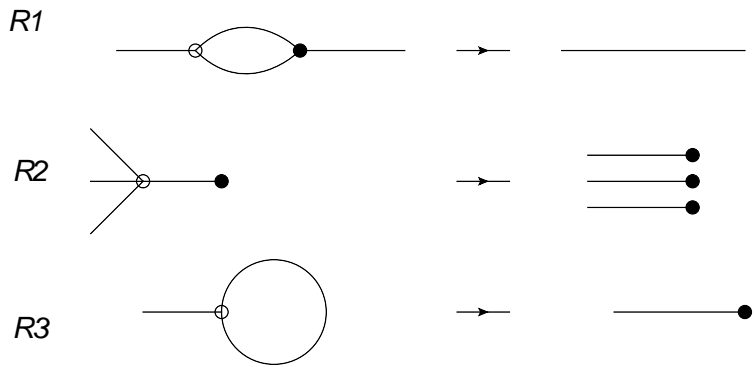}
\caption{Three reductions on a network. R1: bubble reduction. R2: leaf removal. R3: loop removal.
}
\label{netreduction}
\end{figure}

Before discussing the detailed network from the quadrilateral. Let's first define  a special kind of network called minimal network. 
There are three types of reductions which can be used to reduce the face of a network as 
shown in figure.~\ref{netreduction}. R1 is called bubble reduction and R2 is called leaf reduction. R3 is used to remove loop which
is actually a special case of R1: we add a black vertex on the middle of the loop to form a bubble and use bubble reduction to remove it. 
The loop reduction R3 is not applied to the loop winding around the puncture since such loop has non-trivial homology. The definition of the minimal 
network is:

\textbf{Definition}: A minimal network is a planar network for which one can not reduce the face using M1-M3 move plus square move and R1-R3 reductions.

\begin{figure}[htbp]
\small
\centering
\includegraphics[width=8cm]{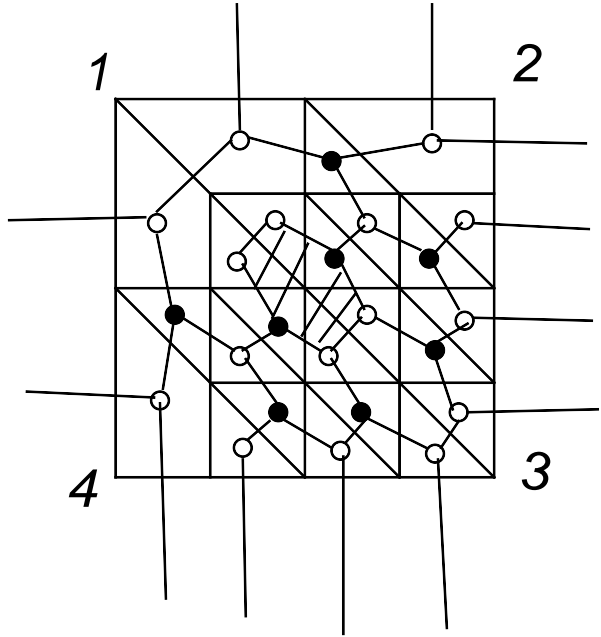}
\caption{The network for a quadrilateral with partitions $[1,1,1,1],[1,1,1,1],[2,2], [2,2]$. The shaded region is 
a bubble and the glued network is not minimal.}
\label{E7}
\end{figure}

The network from three punctured triangle is always minimal from our construction. Similarly,
 the network constructed for the quadrilateral bounded by four full punctures is minimal. However, the network for some non-full puncture case is non-minimal. 
Let's consider a four punctured theory 
 with partitions $[1,1,1,1],[1,1,1,1],[2,2], [2,2]$. The dot diagram and the network is depicted in figure.~\ref{E7}.

It is easy to see that the shaded surface is actually a bubble: two white
vertices with only two edges are removed, and  two black vertices on the boundary of the shaded surface can be contracted 
to form a bubble.  So the naive glued network is not minimal and
one can use the bubble reduction to reduce it.

However,  the network is still not minimal after even removing the obvious bubble. New bubbles may appear after several square moves. 
It is very hard in advance to see whether a network is minimal though and a systematical way of finding the bubbles is needed. 
Such method has indeed been proposed in the literature, see for example \cite{goncharov-2011,postnikov-2006} for more 
details. We will summarize the basic method here.

\begin{figure}[htbp]
\small
\centering
\includegraphics[width=10cm]{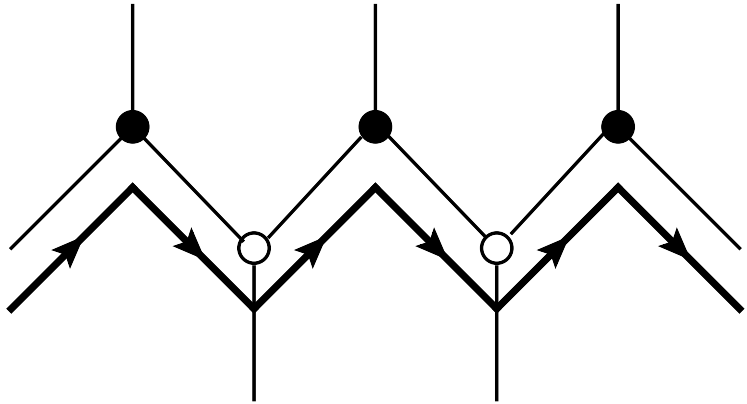}
\caption{A zig-zag path which is turning left at white vertex and turing right at the black vertex.}
\label{zig}
\end{figure}

A special path called zig-zag path is defined on a bipartite graph. Such path is turning left maximally at white vertex and turing right at black vertex. 
See figure.~\ref{zig} for an example. 

\begin{figure}[htbp]
\small
\centering
\includegraphics[width=10cm]{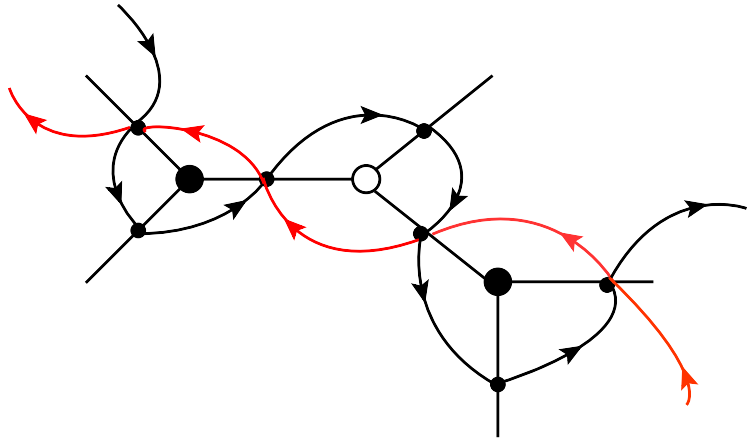}
\caption{Build the strand for a bipartite network.  Put one vertex on each edge and connect these dots around each vertex. The orientation is taken to clockwise for the white vertex and 
counterclockwise direction for the black vertex.}
\label{strand}
\end{figure}

Let's introduce another object called strand  into the network to represent the zig-zag path. Put one vertex on each internal edge and connect the dots around each vertex. Take the 
clockwise orientation for segments around the white vertex and counterclockwise orientation for the black vertex, see figure.~\ref{strand} for the illustration. The network is 
replaced by another network with all four point vertices. A strand is defined as an oriented path such that it represents the zig-zag path.
A piece of strand is also shown in figure.~\ref{strand}.
\begin{figure}[htbp]
\small
\centering
\includegraphics[width=10cm]{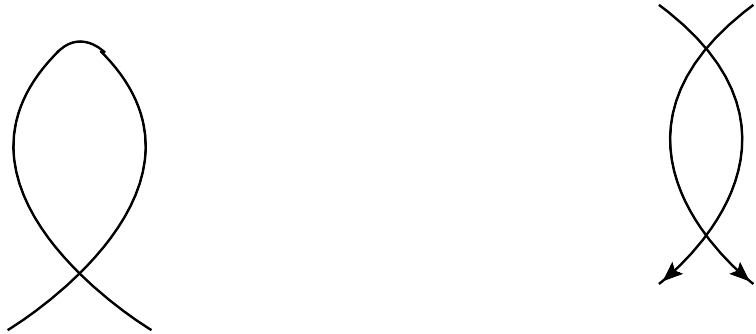}
\caption{The bad configurations for the strand which means that the network is not minimal. }
\label{bad}
\end{figure}

A network is minimal if there is no self-intersections and parallel bigons for the strands crossing as depicted in figure.~\ref{bad}. The network for one triangle is always
a minimal one by our construction since it is defined by removing some edges of a network and reduce it.  The glued network is not minimal if there is type b trapezium in one 
of triangulation of the quadrilateral, because the existence of the type b trapezium reduces the number of surfaces from our construction.

The square move can be done on the minimal network and produce equally good network. Such square move acts on quiver as the quiver mutation.
 In the language of quiver, we are only allowed to do cluster transformation on the quiver nodes which have four arrows attached 
on it. The quiver under such special transformations will  always have 
nice form. On the other hand, the quiver will quickly become a huge mess if  general quiver mutations are allowed. The special fact about the 
SU(2) theory is that all the quiver nodes have four arrows for the regular triangulation and the quiver mutation can be done on any node in arbitrary order.

\begin{figure}[htbp]
\small
\centering
\includegraphics[width=6cm]{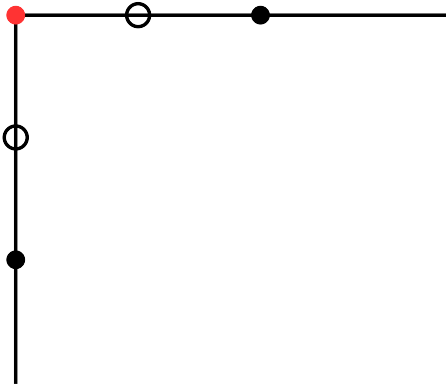}
\caption{The network would be non-minimal if the boundary of the dot diagram has this form at any vertex.}
\label{nonmini}
\end{figure}

The network for the quadrilateral is not minimal if the two adjacent punctures $(Y_1, Y_2)$ have the following form: the height of the first
column of  $Y_1$ and the height of the last column of $Y_2$ is bigger than one, see figure. \ref{nonmini} for the illustration.

\subsubsection{Flip and composite quiver mutation}
Let's focus on the case of puncture configuration where the network is minimal.
There are two different triangulations for the quadrilateral,  these two triangulations are related by the so-called flip.
There would be two minimal networks which should be related by the square move.  We have the following conjecture:

\textbf{Conjecture} 1:  The two minimal networks of two different triangulations of a quadrilateral are related by a sequence of 
square moves.

Let's give a proof of this conjecture by first introducing the concept of permutation from the minimal network on a quadrilateral.
There are $n_r$ number of external lines coming out of the edges representing a puncture with $n_r$ columns.
Let's start from one of the external lines and follow the zig-zag path: turn maximally right at black vertex and maximally left
at white vertex,  the path will reach to another external lines of the network on the other boundary. If we assign numbers to all the boundary 
edges, then each path maps one number to another  and the whole network just defines a permutation. The following theorem
 is proved in \cite{postnikov-2006}.

\textbf{Theorem}: Two minimal network related by square move defines the same permutation.

\begin{figure}[htbp]
\small
\centering
\includegraphics[width=10cm]{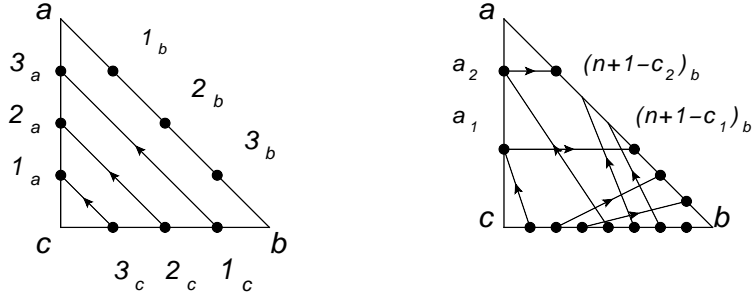}
\caption{Left: The permutation defined by the network on a triangle with three full punctures, the arrows represent the direction of the mapping. 
Right: The permutation defined on a triangle with a generic puncture and two full punctures.}
\label{permutation}
\end{figure}

We can use this theorem to prove that the two minimal networks in the full puncture case define the same permutation and therefore
they are related by square move. The permutation defined by the triangle with three full punctures is easy to find: let's 
replace external lines of the network  by points  on the edge representing puncture $a$  and label them with numbers $(1_a,2_a,...n_a)$ 
in clockwise direction. The numbering and the permutation of SU(3) triangle is shown one the left of figure.~\ref{permutation}. 
The permutation maps $k_c$ to $(n+1-k)_a$, where the subscript letter represents the edge the number is living at and the number is 
our prescribed numbering. 

\begin{figure}[htbp]
\small
\centering
\includegraphics[width=10cm]{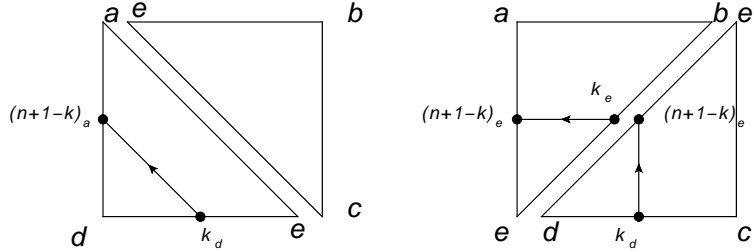}
\caption{Left: The permutation maps points $k_d$ to $(n+1-k)_a$ in this triangulation. 
Right: The point $k_d$ is also mapped to point $(n+1-k)_a$ using the permutation of two triangles.}
\label{permuationful}
\end{figure}

Now let's move to the quadrilateral using the permutation defined on a triangle. Consider an arbitrary point on the boundary representing puncture $d$ with numbering $k_d$, this point is mapped to
$(n-k)_a$ in the triangulation shown on the left of figure.~\ref{permuationful}, where only the permutation of  one triangle is used. 
In another triangulation depicted on the right of figure.~\ref{permuationful}, the point $k_d$ is first mapped to $(n+1-k)_e$ in the triangle bounded by $cde$, the same point 
has another numbering $k_e$ on edge $eb$ according to our convention and map to $(n+1-k)_a$ from the permutation of this second triangle. 
So these two triangulations define the same permutation and the two minimal  
network is related by a sequence of square move  based on the above theorem. A detailed study of the square move sequence will be given later. 

If  one of the puncture is generic with partitions $[n_1,\ldots,n_r]$ and we label the corresponding points
on the boundary $ac$ by numbers $[a_1,\ldots,a_r]$ where $a_i=\sum_{r-i+1}^{r}n_j$,  we label the other two boundaries 
with the same numbering as the full puncture case. By direct observation, we have the following permutation as shown on the right of 
figure.~\ref{permutation}

1. $a_i$ is mapped to $(n+1-a_i)_b$.

2. $(k)_b$ is mapped to $(n+1-k)_c$.

3.  For the dots on edge $bc$, $n_c$ is mapped to $a_1$, $(n-1)_c,\ldots,(n-a_1+1)_c$ is mapped to $((n-a_1+1+1)_b,\ldots,n_b)$. In general, $(n-a_i)_c$ is mapped 
   to $a_{i+1}$; The points between $(n-a_i)_c$ and $(n-a_{i+1})_c$ are mapped to the points between $(n-a_{i+1})_b$ and $(n-a_{i+2})_b$ in an
   ordered  way.
 
 It is now easy to prove that the two triangulations of a quadrilateral with one generic puncture define  the same permutation and therefore the networks are 
 related by square move.  Moreover one can also prove that the square move equivalence of two networks with two generic 
punctures arranged such that they are not adjacent to each other.

In the following, we provide some examples which is in agreement with the above conjecture. The quiver language is used extensively to represent the square move.

\begin{figure}[htbp]
\small
\centering
\includegraphics[width=10cm]{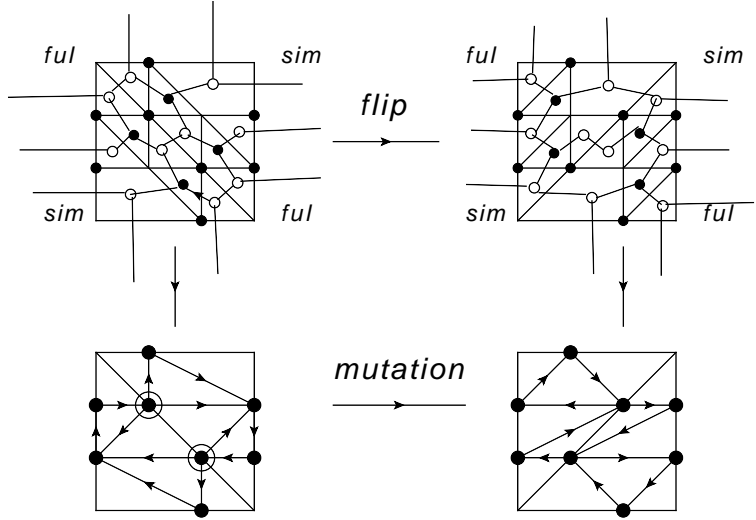}
\caption{Top: The two networks for two triangulations for quadrilateral bounded by two full punctures and two simple punctures of SU(3), the 
cyclic order is indicated too. Bottom: The two quivers for two triangulations are related by quiver mutations. }
\label{mutation1}
\end{figure}

\textbf{Example 1}:  There are two full punctures and two simple punctures of SU(3) theory which are the combinations
 describing  SU(3) theory with six flavors. The punctures are arranged on the quadrilateral 
as indicated in figure.~\ref{mutation1}. The internal puncture is just the full puncture on both triangulation and the dot diagram and 
network are constructed after decorating the internal edge. The networks are minimal and the quiver can be read straightforwardly. The two quivers 
are indeed related by quiver mutations. One just do the quiver mutation on the nodes on the internal edge.
(This can be easily checked using Keller's java program \cite{Keller:2012}). 

\begin{figure}[htbp]
\small
\centering
\includegraphics[width=10cm]{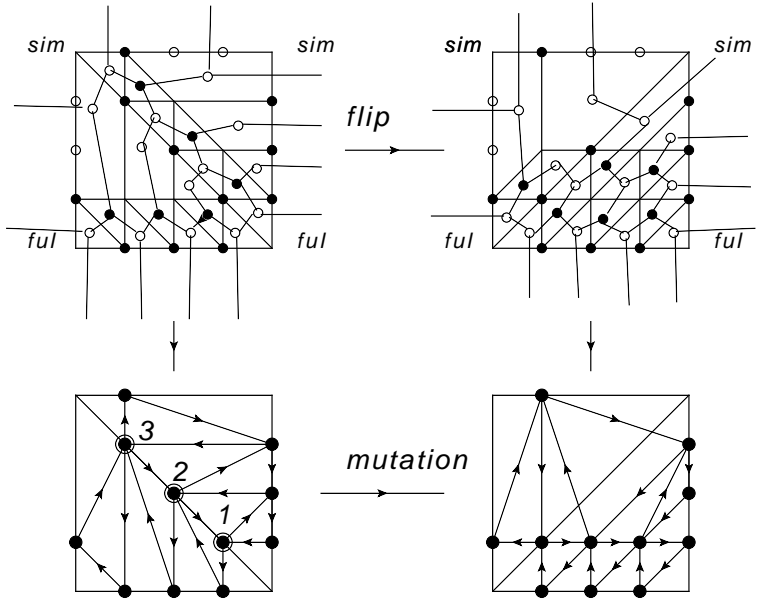}
\caption{Top: The network for quadrilateral of two full punctures and two simple punctures of SU(4) theory. The internal decoration on the left is a full puncture and 
the decoration on the right triangulation is a puncture with partition $[2,1,1]$. Bottom: Two quivers of the network on top are rested by quiver mutation; the order of
mutations are indicated.}
\label{mutation2}
\end{figure}

\textbf{Example} 2:  The decorations on internal edges of two triangulation of this example are different comparing with the first example.
The four punctured theory is SU(4) with eight fundamentals which is represented by two simple punctures and two full punctures.
 The cyclic order is $(sim, sim, ful, ful)$ as shown in
figure.~\ref{mutation2}. The internal puncture is full in one triangulation which corresponds to collide a simple puncture and a full puncture, 
and the network and the quiver is depicted on the left of figure.~\ref{mutation2}.  On the contrary, the internal puncture has partition $[2,1,1]$
in another triangulation in which two simple punctures are colliding. The network and the quiver is depicted on the 
right of figure.~\ref{mutation2}. The two quivers of two triangulations are related using a sequence of quiver mutations which represents
the square move. The order of quiver mutation is also given. 

Let's discuss some general aspects of square move sequence representing the flip. The  full puncture case has a very nice pattern: the squares are all living on the 
internal edge with no arrows between them, we first do square move on those faces with arbitrary order;  then some other
surfaces become squares after doing contractions and there are also no arrows between them, we 
could do the square moves in arbitrary order in second step. We continue doing square moves until no new square appears.  Notice that some surfaces may 
undergo multiple square moves. This is in agreement with the rule given by
Fock and Goncharov \cite{fock-2003}. They describe the sequence using the quiver language. Here we just reproduce what they found using the network and square move.

\begin{figure}[htbp]
\small
\centering
\includegraphics[width=10cm]{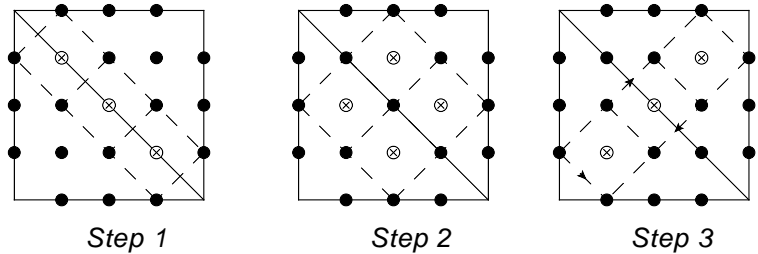}
\caption{Three steps for quiver mutations representing the flip.}
\label{Gmutation1}
\end{figure}

The FG rules is best described using the dot diagram on the quadrilateral in which black dots are the quiver nodes.
The quiver mutations representing flip can be done
in $N-1$ steps. In step $i$, we inscribe a rectangle with lengths $\sqrt{2}(i\times (N-i))$ inside the quadrilateral.
The sides with length $(N-i)\sqrt{2}$ is in parallel with the diagonal direction. Then further decompose the 
rectangle into $\sqrt{2}\times\sqrt{2}$ square, we mutate the quiver nodes at the center of each 
little square at this step, See figure.~\ref{Gmutation1} for the description of  SU(4) theory.
The quiver after these sequence of quiver mutations is the same as the quiver from the quadrilateral derived by flip. 
Notice that in each step the mutated quiver node has four edges which is a square in the network.

The square move sequence is similar for the non-full puncture case if the glued network is minimal.
One still have the same rectangle separation for each step, and we do the square move for 
each square surface whose boundary is completely inside the rectangle.  
After each step, some new squares  appear and  square move will be done 
on those surfaces in next step.

\newpage
\subsubsection{Four puncture theory}
\begin{figure}[htbp]
\small
\centering
\includegraphics[width=10cm]{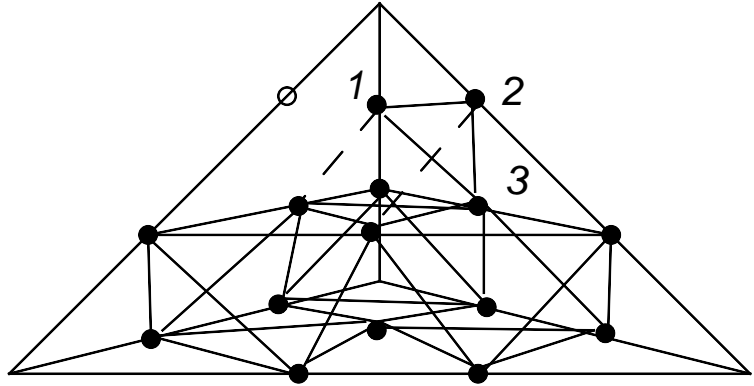}
\caption{The tessellation for the tetrahedron with one simple puncture of SU(3) theory. We project one  small triangle labeled by $123$ on
the surface cut by the simple puncture. }
\label{tetra}
\end{figure}

The full network is derived by gluing the minimal network of two quadrilaterals.  The network is best described by cutting one small 
tetrahedra around each puncture as we have done for the three puncture case. The extra triangles is projected on the surface created by 
this cut, see for example one simple puncture of SU(3) theory in figure.~\ref{tetra}. Then put the appropriate colored vertex on each triangle 
and a network can be easily found.

\begin{figure}[htbp]
\small
\centering
\includegraphics[width=10cm]{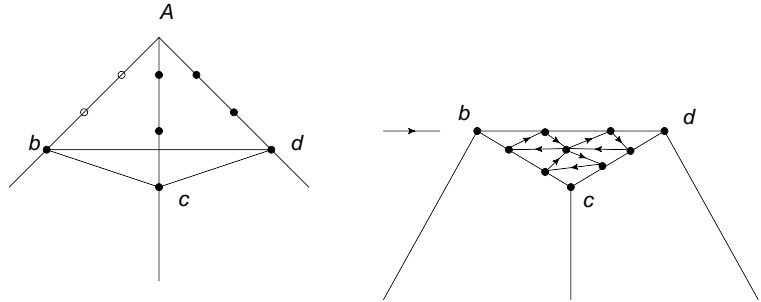}
\caption{The effect of first column of the puncture is to cut a $n_1$ dimensional tetrahedron. }
\label{tetra1}
\end{figure}

The above observation is actually true for any punctures with the Young Tableaux $[n_1, 1,1,1...1]$, the effect of the first column is 
to cut a tetrahedron with height $n_1$, see figure. \ref{tetra1}.  There is a $n_1$ dimensional triangle after the cut and the quiver nodes and arrows on 
this triangle is coming from the triangle $Acd$ in the left part.

The above observation does not depend on the cyclic order of the punctures, which means that the quiver is actually independent of 
the cyclic choices if all the punctures have at most one column with height larger than one. We have the following theorem.

\textbf{Theorem}: The minimal network or the quiver using different cyclic paths
 are the same if all the punctures have at most one column with height larger than one.

\begin{figure}[htbp]
\small
\centering
\includegraphics[width=10cm]{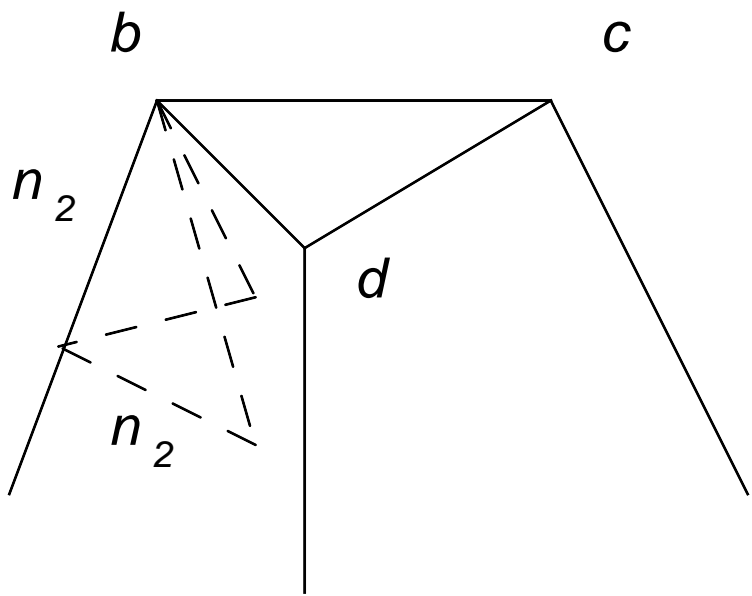}
\caption{The effect of second column of the puncture is to cut a  $n_2$ dimensional tetrahedron. }
\label{secondcut}
\end{figure}

If the puncture has the form $[n_1,n_2,...]$ with $n_2>1$, then the effect of the second column is to cut another $n_2$ dimensional 
tetrahedron on one edges, see figure. \ref{secondcut}. However, in this case there are choices since one can cut it on any
edges ending on $A$, this choice is equivalent to the choice of the cyclic paths for the four puncture.

\newpage

\subsection{General case}

\subsubsection{Sphere with arbitrary number of punctures}
The generalization to the sphere with $n$ punctures is straightforward. First consider
a regular triangulation of the punctured sphere, it is always possible to find a loop connecting all
these punctures.  Choose any of the closed loop connecting all the punctures and decorate the 
edges from the information of the punctures. The decoration on the internal edge is determined by 
the formula derived by the S-duality study of the gauge theory.

\begin{figure}[htbp]
\small
\centering
\includegraphics[width=10cm]{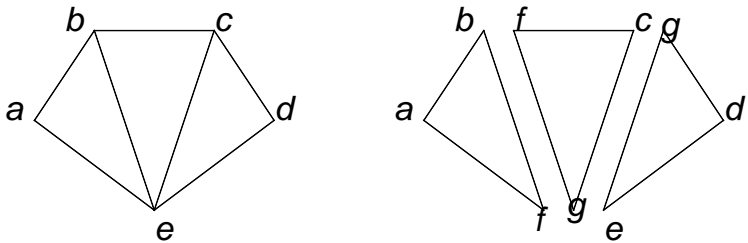}
\caption{The decomposition of the pentagon into triangles.}
\label{pentag}
\end{figure}

Each polygon can be decomposed into several triangles.  Let's take the pentagon for simplicity. 
The pentagon is decomposed into three triangles as shown in
figure.~\ref{pentag}.  There are two new punctures appearing in this decomposition. The puncture 
$g$ and $f$ are determined as the following
\begin{equation}
f_i=min(a+b, c+d+e, i-1),~~g_i=min(a+b+c, e+d, i-1).
\end{equation}

The decoration of the internal edges are completely fixed by the boundary edges. There are maybe  four punctured 
theory without dimension $N$ operators and we need to study the flip for these special cases. For example, if punctures $a,b,c$ are simple punctures while the 
puncture $d$ and $e$ are full for the SU(3) theory. The four punctured theory bounded by punctures $abcg$ has no dimension 3 operators but the glued network on the quadrilateral
can be constructed without any problem.  One could still show that the flip of the triangulations are represented by 
the square move if the local network of the quadrilateral satisfies the minimality condition, etc.

Let's now study some more detail on one puncture whose Young Tableaux has the form $[n_1,1,1,...,1]$.  Let's assume there are 
$n$ edges ending on this puncture in the triangulation, and according to our prescription, one of the edge is decorated 
according to our rule, namely, there are white dots on this edge. There are only $(n-2)$ $n_1\times n_1$  triangles filled with
quiver nodes and quiver triangles. Now we could project the quiver nodes and arrows of those triangles into 
the surface defined by the polygon in figure. \ref{cut2} as one could check, so  the effect of the non-full puncture of this type
is to cut a polyhedron around the puncture. Notice that this observation does not depend on the cyclic choice 
of the paths, therefore we have the following statement:

\textbf{Theorem}: The quiver does not depend on the cyclic choice of the paths if all the punctures have at most 
one column with height bigger than one.

\begin{figure}[htbp]
\small
\centering
\includegraphics[width=10cm]{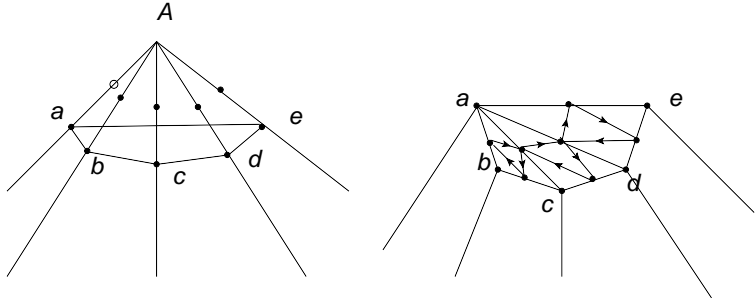}
\caption{The effect of a non-full puncture is to cut a polyhedron around the puncture. }
\label{cut3}
\end{figure}

If the Young Tableaux has the form $[n_1, n_2, ...]$, the effect of second column is to cut a tetrahedron around one 
of the vertices created in the first cut due to $n_1$. However, different choices do give different quivers.

\newpage

\subsubsection{Higher genus surface}
The construction can be easily generalized to  theories of class ${\cal S}$defined using higher genus Riemann surface. Consider a genus one Riemann surface with 
one puncture, geometrically, such Riemann surface is derived by gluing two legs of the three punctured sphere.
Physically, such gluing corresponds to glue the diagonal of two SU(N) flavor group. Notice that it is only possible to 
form a handle using two full punctures.  

\begin{figure}[htbp]
\small
\centering
\includegraphics[width=10cm]{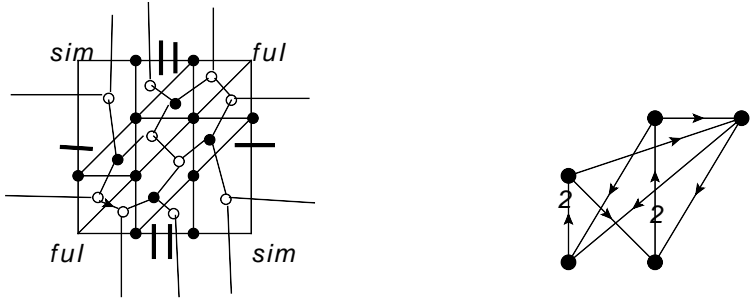}
\caption{The triangulation of one punctured torus. The puncture is a simple puncture of SU(3) theory and the quiver is read from the network.}
\label{torus}
\end{figure}

The triangulation of the one punctured torus is depicted in figure.~\ref{torus} in which  the simple puncture of SU(3) theory is used.
The boundaries of the quadrilateral are identified as shown in the figure.~\ref{torus}. The decoration of the diagonal edge is always the 
full puncture. 

It is straightforward now to find a network on this triangulation and the glued network is minimal. The flip on the diagonal edge
corresponds to mutating two quiver nodes on the diagonal edge. Amazingly the new quiver after mutation has the same form as 
the original quiver with all the arrows reversed which is exactly like what happens for the SU(2) theory. Such properties are 
true for any other punctures.

There is no problem finding a network for any higher genus Riemann surface with any combinations of the punctures.  The construction is applicable
even if there is no puncture. For example, a genus two Riemann surface without any puncture can be formed by gauging the punctures of 
two genus one Riemann surface with a full puncture. At the level of the quiver, one double the quiver node on the diagonal edge of the 
torus triangulation and gluing two quivers together by identifying two sets of these nodes.

\section{Conclusion}
 A combinatorial method of building a network and a quiver is developed for any  theories of class ${\cal S}$ in this paper.
 The quiver has the correct dimensions as the framed moduli space of the flat connections. 
 We conjecture that the corresponding cluster coordinates are the generalization
 of FG coordinates constructed in \cite{fock-2003} for the full puncture.  The constructions are based on a triangulation 
and a cyclic path from the triangulation on the punctured Riemann surface. 
Different triangulations with fixed cyclic boundary are connected by a sequence of local flip operations .
We prove that the quivers for different triangulations are related by a sequence of 
 quiver mutations corresponding to square move if there is no bad corner in the cyclic choice of the punctures. 
 We also prove that the quiver for different choices of the cyclic order are the same if all the punctures have at most one column with height larger than one. 
 One of possible interpretation for the dependence on the cyclic choice is that they define different framed local system, we
 would like to return to this in the near future.

 We do not consider the cluster coordinates in any detail in this paper. To really prove that our construction gives the 
 FG coordinates for moduli space of flat connections, we need to reconstruct holonomy around the puncture from
 these cluster coordinates and show that the holonomy have  the correct form dictated by the Young Tableaux. The method of 
 calculating the holonomy matrix in the full puncture case is given in \cite{fock-2003}. Currently we are trying to generalize their method to
 the non-full puncture case. 
 
 There are various generalizations one could consider. First, we only consider  theories of class ${\cal S}$i which require regular puncture, and the generalization
 to asymptotical free (AF) theories  and Argyres-Douglas (AD) theories is given in a sequel to this paper. The six dimensional 
 construction for some of the AF theories are given in \cite{Nanopoulos:2010zb} and the second order irregular punctures are needed. The AD theories 
 involves higher order irregular singularities and a complete classification will be given in \cite{Xie:2012ff}. The existence of irregular
 singularity introduces boundary with marked points to the Riemann surface.The triangulation and the network is actually already given
 here if one could find the Young-Tableaux sequences for the irregular singularities. We will develop this classification first \cite{Xie:2012ff} before
coming to the cluster coordinates for these theories \cite{Xie:2012mm}. Second, one hope to generalize the same construction to the theories
  derived using $D_n$ \cite{Tachikawa:2009rb} and $E_n$ six dimensional $(2,0)$ theory. Third, there are many other $\mathcal{N}=2$ theories which do not have a 
  six dimensional construction but can be engineered using toric geometry \cite{Katz:1997eq}; A dimer model which is  a bipartite network on a torus 
  can be described once the newton polygon of the toric geometry is given. In fact, the brane web constructions considered in this 
  paper has close relationship with the dimer model as shown in \cite{Franco:2005rj}. Naively, we would expect that the corresponding quiver from the dimer is the 
  right one to describe the BPS spectrum of these theories. It would be  interesting
  to verify whether the above picture is right or not.

There are many applications of these coordinates. One could use the method developed in \cite{Gaiotto:2008cd,Gaiotto:2009hg, Gaiotto:2010be, Gaiotto:2011tf, Cecotti:2011rv, Alim:2011ae, Alim:2011kw} to find the stable BPS spectrum and 
the wall crossing behavior of the corresponding four dimensional theories. We outline how these methods can be used for the higher
rank theory and leave the details to further publication. First, one would like to use the quiver mutation method: our constructions automatically give many sequence of 
quiver mutations which mutate the quiver back to itself, typical example is the pentagon relations.
Such sequence of quiver mutations is essential to find a dilogrithm identifies and the BPS spectrum. 
Second, one can use the representation theory of quiver with superpotential. The superpotential of  the quiver appearing in SU(2) theory is found in \cite{labardinifragoso-2008} 
which is used to  to study wall crossing in \cite{Alim:2011ae,Alim:2011kw}. The generalization to the higher 
rank theory seems to be obvious: there is a superpotential term for each black dot on the minimal network and a term for any closed zig-zag loops 
winding around the puncture. Simple examples show that this prescription is consistent with flip.  Third, the cluster coordinates are used to 
classify the line operators of the gauge theory \cite{Drukker:2009tz}, it would be interesting to do the  same classification using our cluster coordinates to the higher
rank theory. Fourth, once the holonomy construction is given, one can express the expectation value of line operators in terms of the cluster coordinates, and it is possible to 
use the method developed in \cite{Gaiotto:2010be, Gaiotto:2011tf} to study the wall crossing in the presence of line defects and surface defects.

One could also use these coordinates to study the higher rank complex Chern-Simons theory on a three hyperbolic manifold \cite{Dimofte:2011ju} and 
construct the corresponding $N=2$ gauge theory \cite{Dimofte:2012}.  The generalization of  method used in  \cite{Cecotti:2011iy} to study 
three dimensional theory seems to be also possible using these cluster coordinates, i.e. the simplest case is the pentagon with five full punctures which represents
a higher rank analog of the AD theories,  the BPS spectrum of this theory can be found easily using the quiver mutation sequence, it is interesting to 
carry a detailed study following the approach of \cite{Cecotti:2011iy}. 

A quantum version of the cluster algebra is also developed in \cite{fock-2007}. The Poisson structure in the cluster coordinates is extremely simple and the 
quantization is relatively easy. The use of cluster coordinates to the quantum teichmuller theory is given by \cite{chekhov-1999-120} and it is conjectured that 
Hilbert space of the quantum Teichmuller theory is same as the Liouville theory. We expect that the same story is true for the higher Teichmuller theory, 
i.e. the quantum higher Teichmuller theory is isomorphic to the Toda field theory. And it is possible to get the conformal block of Toda field theory using these 
nice coordinates and get the expectation value of the line operators and surface operators based on AGT conjecture \cite{Drukker:2009id,Alday:2009fs,Drukker:2010jp}.

Goncharov and Kenyon associated an integrable system to a dimer model on torus \cite{goncharov-2011}. Our network is very similar to the dimer model and it is also possible to associate 
an integrable system following their approach. This integrable system may be identified with the Seiberg-Witten integrable system and we can use the quantum 
cluster algebra to quantize it. Further, one can construct the Darboux coordinates as described in \cite{Nekrasov:2011bc}  using the cluster coordinates too. These 
connections to integrable system is fun to study.

Finally, these cluster coordinates of Riemann surface with one irregular singularity
are playing important role in studying scattering amplitude of strongly coupled $\mathcal{N}=4$ theory \cite{Alday:2009dv}. In fact, the cluster $y$ coordinates  and its cluster 
transformations are exactly the same as the TBA equation \cite{Fomin_y-systemsand} which is used to calculate the  $\mathcal{N}=4$ scattering amplitude \cite{Alday:2009dv} in the strong coupling limit.
We believe that the general cluster coordinates considered in this paper
might be  useful in studying other quantities like form factor \cite{Maldacena:2010kp}, correlation function, etc. The same mathematical structure is also 
used amazingly in studying leading singularity of weakly coupled $\mathcal{N}=4$ planar theory and it is natural to expect the constructions considered
in this paper might be useful in that context too.

\begin{flushleft}
\textbf{Acknowledgments}
\end{flushleft}
We would like to thank Murad Alim, Tudor Dimofte, Davide Gaiotto, Yu-tin Huang, Cumrun Vafa, Masahito Yamazaki,  Peng Zhao and especially Nima Arkani-Hamed for useful
discussions.  The research of DX is support by the friends of Institute for Advanced Study and 
acknowledges support by the U.S. Department of Energy, grant DE-FG02-90ER40542.

\bibliographystyle{utphys} 
 \bibliography{PLforRS}

\end{document}